\documentclass[aps,prd,twocolumn,amsmath,superscriptaddress,amssymb,showpacs,floatfix,nofootinbib,longbibliography]{revtex4-1}

\usepackage{graphicx}
\usepackage{dcolumn}
\usepackage{xcolor}
\usepackage{bm}
\usepackage{natbib}
\usepackage{calligra}
\usepackage[T1]{fontenc}
\usepackage{egothic}
\usepackage[T1]{fontenc}
\newfont{\rsfsten}{rsfs10 scaled 1200}
\newfont{\rsfsseven}{rsfs10 scaled 1200}
\newfont{\rsfsfive}{rsfs10 scaled 1200}
\usepackage{epsfig}
\usepackage{units}
\usepackage[utf8]{inputenc}

\newcommand{\be}{\begin{equation}}
\newcommand{\ee}{\end{equation}}
\newcommand{\bea}{\begin{eqnarray}}
\newcommand{\eea}{\end{eqnarray}}

\newcommand{\ie}{{\it i.e.~}}

\newcommand{\eg}{{\it e.g.~}}

\def\lsim{\mathrel{\raise.3ex\hbox{$<$\kern-.75em\lower1ex\hbox{$\sim$}}}}
\def\gsim{\mathrel{\raise.3ex\hbox{$>$\kern-.75em\lower1ex\hbox{$\sim$}}}}

\begin{document}

\vskip 0.2in
\hspace{13cm}\parbox{5cm}{FERMILAB-PUB-19-091-A}
\hspace{13cm}
\vspace{0.6cm}

\title{A Robust Excess in the Cosmic-Ray Antiproton Spectrum:  Implications for Annihilating Dark Matter}


\author{Ilias Cholis}
\email{cholis@oakland.edu, ORCID: orcid.org/0000-0002-3805-6478}
\affiliation{Department of Physics, Oakland University, Rochester, Michigan, 48309, USA}
\author{Tim Linden}
\email{linden.70@osu.edu, ORCID: orcid.org/0000-0001-9888-0971}
\affiliation{Center for Cosmology and AstroParticle Physics (CCAPP) and Department of Physics, The Ohio State University Columbus, Ohio, 43210 USA}
\author{Dan Hooper}
\email{dhooper@fnal.gov,  ORCID: orcid.org/0000-0001-8837-4127}
\affiliation{Fermi National Accelerator Laboratory, Center for Particle Astrophysics, Batavia, Illinois, 60510, USA}
\affiliation{University of Chicago, Department of Astronomy and Astrophysics, Chicago, Illinois, 60637, USA}

\date{\today}

\begin{abstract}

An excess of $\sim$10-20 GeV cosmic-ray antiprotons has been identified in the spectrum reported by the \textit{AMS-02} Collaboration.  The systematic uncertainties associated with this signal, however, have made it difficult to interpret these results. In this paper, we revisit the uncertainties associated with the time, charge and energy-dependent effects of solar modulation, the antiproton production cross section, and interstellar cosmic-ray propagation. After accounting for these uncertainties, we confirm the presence of a 4.7$\sigma$ antiproton excess, consistent with that arising from a $m_{\chi} \approx 64-88$ GeV dark matter particle annihilating to $b\bar{b}$ with a cross section of $\sigma v  \simeq (0.8-5.2) \times 10^{-26}$ cm$^{3}$/s. If we allow for the stochastic acceleration of secondary antiprotons in supernova remnants, the data continues to favor a similar range of dark matter models ($m_{\chi}\approx 46-94$ GeV, $\sigma v \approx (0.7-3.8)\times 10^{-26}$ cm$^3/$s) with a significance of 3.3$\sigma$. The same range of dark matter models that are favored to explain the antiproton excess can also accommodate the excess of GeV-scale gamma rays observed from the Galactic Center.


\end{abstract}

\maketitle

\section{Introduction}
\label{sec:introduction}

Measurements of antimatter in the cosmic-ray spectrum have long been used to advance 
our understanding of high-energy phenomena in the Galaxy~\cite{Bergstrom:1999jc, Hooper:2003ad, Profumo:2004ty, Bringmann:2006im, 
Pato:2010ih} . To this end, \textit{AMS-02} has measured the cosmic-ray antiproton spectrum and antiproton-to-proton ratio~\cite{Aguilar:2016kjl}, and are searching for cosmic-ray antimatter nuclei~\cite{AMSLaPalma}. Measurements such as these provide a powerful way to search for new physics, including the annihilation or decay of dark mater particles in the halo of the Milky Way. 

Over the past several years, a number of groups have reported the presence of an excess of $\sim$10-20 GeV antiprotons, consistent with the annihilation products of dark matter~\cite{Hooper:2014ysa, Cirelli:2014lwa, Bringmann:2014lpa, Cuoco:2016eej, 
Cui:2016ppb}. Moreover, an excess of GeV-scale gamma-rays from the Inner Galaxy has now been firmly confirmed~\cite{Hooper:2010mq, 
Hooper:2011ti, Abazajian:2012pn, Gordon:2013vta, Daylan:2014rsa,Calore:2014xka,TheFermi-LAT:2015kwa}. Although these two signals are sensitive to very different systematic uncertainties and backgrounds, it is intriguing that they could both be explained by a $\sim$60 GeV dark matter particle with an annihilation cross section near that predicted for a generic thermal relic~\cite{Daylan:2014rsa, Calore:2014nla, Agrawal:2014oha, Caron:2015wda,Cuoco:2017rxb}. The concordance between these two signals is suggestive and provides considerable motivation for additional indirect searches for dark matter (see e.g. Refs.~\cite{Geringer-Sameth:2014qqa,Fermi-LAT:2016uux}). 

Taken at face value, the statistical significance of the \textit{AMS-02} antiproton excess is quite high. The authors of Ref.~\cite{Cuoco:2016eej}, for example, assess the significance of this signal to be $\sim$\,4.5$\sigma$. It is somewhat challenging, however, to quantify the systematic uncertainties associated with this excess, including, i) uncertainties associated with the injected spectra of cosmic-ray protons, helium, and heavier nuclei, ii) uncertainties associated with the propagation of cosmic rays through the interstellar medium (ISM), iii) uncertainties associated with the antiproton production cross sections for proton-proton, proton-nucleon and nucleon-nucleon inelastic collisions, and iv) uncertainties associated with the impact of the solar wind on the cosmic-ray spectra observed at Earth~\cite{Cholis:2017qlb}.

\begin{table*}[t]
    \begin{tabular}{ccccccccc}
         \hline
           ISM Model & $\delta$ & $z_{L} ({\rm kpc})$ & $D_{0} \times 10^{28}$ (cm$^2$/s) & $v_{A}$ (km/s) & $dv_{c}/d|z|$ (km/s/kpc) 
           & $\alpha_{1}$ & $\alpha_{2}$ & $R_{\rm br}$ (GV) \\
            \hline \hline
            I &  0.40 & 5.6 & 4.85 & 24.0 & 1.0 & 1.88 & 2.38 & 11.7 \\      
            II &  0.50 & 6.0 & 3.10 & 23.0 & 9.0 & 1.88 & 2.45 & 11.7 \\
            III &  0.40 & 3.0 & 2.67 & 22.0 & 3.0 & 1.87 & 2.41 & 11.7 \\
            \hline \hline 
        \end{tabular}
\caption{The parameters for the cosmic-ray injection and propagation models used in this study. Each of these models provides a good overall fit to the observed cosmic-ray proton, helium, carbon, and boron-to-carbon ratio spectra up to 200 GV~\cite{Cholis:2015gna, Cholis:2017qlb}. These ISM parameters are not picked in advance to fit the $\bar{p}/p$ ratio.} 
\label{tab:ISMBack}
\end{table*}

The remainder of this article is structured as follows. In Sec.~\ref{sec:method} 
we describe our treatment of the systematic uncertainties listed in the previous paragraph. 
We then present our main results in Sec.~\ref{sec:results}, finding that the antiproton-to-proton ratio measured by \textit{AMS-02} favors the presence of a contribution from annihilating dark matter at the level of 4.7$\sigma$. In Sec.~\ref{sec:SSA}, we discuss how the stochastic acceleration of antiprotons in supernova remnants can impact our results, favoring a similar range of dark matter models but with a somewhat lower statistical significance of 3.3$\sigma$. The inclusion of this contribution also leads to a better fit to the antiproton spectrum at energies above $\sim$100 GeV. In Sec.~\ref{sec:conclusions} we summarize our results and discuss them within the larger context of indirect searches for annihilating dark matter.

\section{Methodology and Assumptions}
\label{sec:method}

In this section, we describe our efforts to quantify the systematic uncertainties associated with the cosmic-ray antiproton spectrum. Readers interested only in the results of our analysis may feel free to skip to Sec.~\ref{sec:results}.

\subsection{Cosmic-Ray Injection and Propagation in 
the Interstellar Medium}
\label{sec:ISM}

Antiprotons can be produced as secondary cosmic rays when energetic cosmic-ray primaries (\ie cosmic rays accelerated by supernova remnants) collide with interstellar gas. Cosmic rays that acquire their energy through first order Fermi acceleration are generally expected to exhibit a power-law spectrum, $dN/dE \propto E^{-\alpha}$, with a
typical spectral index of $\alpha \sim 2.0$ (see \eg Ref.~\cite{Mertsch:2018bqd}). A large number of supernova remnants contribute to the cosmic-ray spectrum, with a variety of ages and at a range of distances from the Solar System. In our calculations, we adopt the following parameterization for the average injected spectra of cosmic-ray protons and nuclei:
\begin{eqnarray}
 dN/dR \propto \begin{cases} R^{-\alpha_{1}}, \,\,\,\,\,\,\,
     \textrm{for} \; R<R_{\rm br}\\  
     R^{-\alpha_{2}}, \,\,\,\,\,\,\, \textrm{for} \; R>R_{\rm br},
     \end{cases}
\label{eq:Inj}
\end{eqnarray}
where $R$ is the cosmic-ray rigidity (the momentum-to-charge ratio). For simplicity, we adopt the same values of $\alpha_1$, $\alpha_2$ and $E_{\rm br}$ for protons, helium and other nuclear species, and do not account for the spectral hardening that has been observed at rigidities above $\simeq$\,200 GV in proton, He, Li, Be, B, C, and O cosmic rays~\cite{Aguilar:2015ooa, 
Aguilar:2015ctt, Aguilar:2017hno, Aguilar:2018njt}. While ignoring the possibility of a spectral break above 200 GV could impact the antiproton 
flux by $\simeq 1\%$ at 1 GV (2\%, 20\% at 10 and 100 GV, respectively), such a feature cannot produce any spectral features in the $\sim$10-20 GeV range and has a negligible impact 
on our results. Furthermore, given that the B/C ratio shows no evidence of such hardening~\cite{Aguilar:2016vqr}, we conclude that this high-energy spectral feature is likely to be the result of variations in the local source distribution, not unlike that observed in the spectrum of cosmic-ray positrons produced by pulsars (see e.g. Refs.~\cite{Cholis:2017ccs,Cholis:2018izy}). 

Once injected into the ISM, cosmic rays undergo diffusion, convection, and diffusive reacceleration (more massive nuclei may also experience fragmentation). Energy losses for cosmic-ray nuclei by ionization and Coulomb collisions are also included but have a subdominant impact on our results. We take each of these processes into account by solving the cosmic-ray transport equation numerically, using the publicly available code \texttt{Galprop} v54 1.984~\cite{GALPROPSitev54, GALPROPSite, Strong:2015zva, galprop}.

To model the effects of isotropic and homogenous transport throughout a zone extending up to a half-height of $z_L$ from the Galactic Disk, we adopt the following diffusion coefficient:
\begin{equation}
D_{xx}(R) = \beta D_{0} (R/4 \,{\rm GV})^{\delta},
\label{eq:Diff}
\end{equation}
where $\beta \equiv v/c$ and $\delta$ is the diffusion index associated to the spectral index of magnetohydrodynamic turbulence in the ISM. 
Typical values for $\delta$ are 0.33 for Kolmogorov turbulence~\cite{1941DoSSR..30..301K} 
and 0.5 for the Kraichnan case~\cite{1980RPPh...43..547K}. Values within this range are generally
consistent with the existing body of cosmic-ray data (see e.g. Ref.~\cite{Trotta:2010mx}). 

Diffusive reacceleration is described by a diffusion coefficient in momentum space~\cite{1994ApJ...431..705S},
\begin{equation}
D_{pp} \propto \frac{R^{2}v_{A}^{2}}{D_{xx}(R)},
\label{eq:DiffReAcc}
\end{equation}
where the Alfv$\acute{\textrm{e}}$n speed, $v_{A}$, is the speed at which 
hydromagnetic waves propagate in the ISM plasma. 

The convective wind speed, $v_{c}$, has a gradient perpendicular to the Galactic 
Plane,
\begin{equation}
v_{c} = \frac{dv_{c}}{d|z|} |z|.
\label{eq:Conv}
\end{equation}

To constrain the above parameters, we follow the procedure described in Refs.~\cite{Cholis:2015gna, Cholis:2017qlb},
using data from \textit{AMS-02}, \textit{PAMELA} and \textit{Voyager 1}. Previous studies have found the $\sim$10-20 GeV antiproton excess to be robust to variations in the values of these parameters~\cite{Cuoco:2016eej,Cui:2016ppb,Hooper:2014ysa}. Instead of repeating the same procedure here, we chose to adopt three representative models for cosmic-ray injection and transport (see Table~\ref{tab:ISMBack}), each of which provides a good overall fit to the observed cosmic-ray proton, helium, carbon, and boron-to-carbon ratio spectra up to 200 GV (see also Table I of Ref.~\cite{Cholis:2017qlb}).\footnote{Our ISM models "I", "II" and "III" are the same as ISM models "C", "E and "F" from our previous work in Ref.~\cite{Cholis:2015gna, Cholis:2017qlb}. We find that if we add the ISM models "A", "B" and "D" from Ref.~\cite{Cholis:2015gna}, our results regarding the GeV $\bar{p}/p$ excess at the 5-20 GeV in $E_{\textrm{kin}}$ fall within the same range of significance.}

\subsection{The Antiproton Production Cross Section}
\label{sec:CrossSecPBAR}

The production of antiprotons in the inelastic collisions of high-energy nuclei
has been studied at a number of collider experiments~\cite{Dekkers:1965zz, 
Capiluppi:1974rt, Allaby:1970jt, Guettler:1976ce, Johnson:1977qx, 
Antreasyan:1978cw, Arsene:2007jd, Anticic:2009wd, Acharya:2017fvb}, and this information has been parameterized in several different forms~\cite{Tan:1982nc, 
Tan:1983de, Duperray:2003bd,diMauro:2014zea, 
Kappl:2014hha, Kachelriess:2015wpa, Bernstein:2018ndj}. Most of the 
existing work on this subject has focused on the direct productions of antiprotons in proton-proton 
collisions. Also relevant to the problem at hand is the production of antineutrons, which lead to the production of antiprotons through their decay in the ISM. Furthermore, significant uncertainties apply to the rate of antiproton production from helium and other nuclei, which are collectively responsible for approximately 40\% of the overall flux. 

Using  \texttt{Galprop}, we account for the production of antiprotons from
all cosmic-ray species~\cite{Moskalenko:2001ya}. Based on Ref.~\cite{diMauro:2014zea},
the 3$\sigma$ uncertainty on $\sigma_{pp \rightarrow X+\bar{p}}$  is about 
$\simeq 40 \%$ for antiprotons with a kinetic energy of 1 GeV. This uncertainty is energy dependent, as shown in Fig.~8 of 
Ref.~\cite{diMauro:2014zea}. Following up on our previous work~\cite{Cholis:2017qlb}, we first evaluate the antiproton flux for a given cosmic-ray transport model and cross section, and then marginalize over a flat prior within the energy-dependent
$3\sigma$ uncertainties on the antiproton production cross section, as quoted in Ref.~\cite{diMauro:2014zea}. We do this 
through the following energy-dependent scaling factor:
\begin{eqnarray}
N_{CS}(E_{\textrm{kin}}) = a + b \, \ln \bigg(\frac{E_{\textrm{kin}}}{\rm GeV}\bigg) + c \, \bigg[\ln\bigg(\frac{E_{\textrm{kin}}}{\rm GeV}\bigg)\bigg]^{2}. 
\label{eq:CSpbar}
\end{eqnarray}
The parameters $a$, $b$ and $c$ and are allowed to vary over a large range of values in order to cover the range quoted in Ref.~\cite{diMauro:2014zea}. By allowing this energy dependent cross section to vary without penalty within the quoted 3$\sigma$ uncertainties, we are conservatively allowing for a generous range of behavior in our analysis. By including a greater degree of flexibility in this parameterization (\eg a term proportional to $[\ln(E_{\rm kin}/{\rm GeV})]^3$), one could absorb much of the $\sim$10-20 GeV excess observed in the cosmic-ray antiproton spectrum. We emphasize, however, that the terms contained in Eq.~\ref{eq:CSpbar} more than adequately encompass the physically plausible range of uncertainties associated with this quantity. Moving forward, high precision laboratory measurements of the antiproton production cross section could reduce these uncertainties and substantially increase our ability to search for dark matter annihilation products in the cosmic-ray spectrum.

\subsection{Solar Modulation}
\label{sec:SolMod}

As cosmic rays enter the Solar System, their spectra are modulated by the solar wind and its embedded magnetic field. We use the standard formula to model the impact of the modulation potential~\cite{1968ApJ...154.1011G}:
\begin{eqnarray}
\frac{dN^{\oplus}}{dE_{\textrm{kin}}} (E_{\textrm{kin}}) &=& \frac{(E_{\textrm{kin}}+m)^{2} 
-m^{2}}{(E_{\textrm{kin}}+m+|Z| e \Phi)^{2} -m^2} \nonumber \\ 
&\times& \, \frac{dN^{\rm ISM}}{dE_{\rm kin}^{\rm ISM}} (E_{\rm kin}+|Z| e \Phi),
\label{eq:Mod}
\end{eqnarray}
where $E_{\textrm{kin}}$ is the kinetic energy of the cosmic ray measured at Earth, Z$e$ and $m$ are the charge and mass of the cosmic ray, $dN^{\oplus}/dE_{\textrm{kin}}$ is the spectrum measured at Earth and $dN^{\rm ISM}/dE_{\rm kin}^{\rm ISM}$ is the spectrum present in the ISM, prior to the effects of solar modulation. Based on Ref.~\cite{Cholis:2015gna}, we adopt the following rigidity, time and charge-dependent modulation potential:
\begin{eqnarray}
\label{eq:ModPot}
\Phi(R,t,q) &=& \phi_{0} \, \bigg( \frac{|B_{\rm tot}(t)|}{4\, {\rm nT}}\bigg) + \phi_{1} \, N'(q) 
H(-qA(t)) \\ 
&\times& \bigg( \frac{|B_{\rm tot}(t)|}{4\,  {\rm nT}}\bigg) \, \bigg(\frac{1+(R/R_0)^2}{\beta (R/R_{0})^3}\bigg) 
\, \bigg( \frac{\alpha(t)}{\pi/2} \bigg)^{4}, \nonumber
\end{eqnarray}
where $B_{\rm tot}(t)$ is the strength of the heliospheric magnetic field at Earth (as measured by \textit{ACE}~\cite{ACESite}), $A(t)$ is its polarity, and $\alpha(t)$ is the tilt angle of the heliospheric current sheet (based on models provided by the Wilcox Solar Observatory~\cite{WSOSite}). $R$ is the rigidity of the cosmic ray prior to entering the Solar System, and $R_0 \equiv 0.5$ GV.

To model the uncertainties associated with solar modulation, we allow for $0.32 \leq \phi_{0} \leq 
0.38$ GV and $0 \leq \phi_{1} \leq 16$ GV, each of which represent the 2$\sigma$ range presented in Ref.~\cite{Cholis:2017qlb}. The quantity $N'(q)$, along with averaged values of $B_{\rm tot}(t)$ and $\alpha(t)$, are given 
in Table II of Ref.~\cite{Cholis:2017qlb} for each six-month interval. 
We perform the Solar modulation correction over each of the these six-month intervals and
calculate the averaged spectra before comparing to the data (see Refs.~\cite{Cholis:2015gna, 
Cholis:2017qlb} for further details).

\begin{figure*}
\hspace{-0.2176in}
\includegraphics[width=2.58in,angle=0]{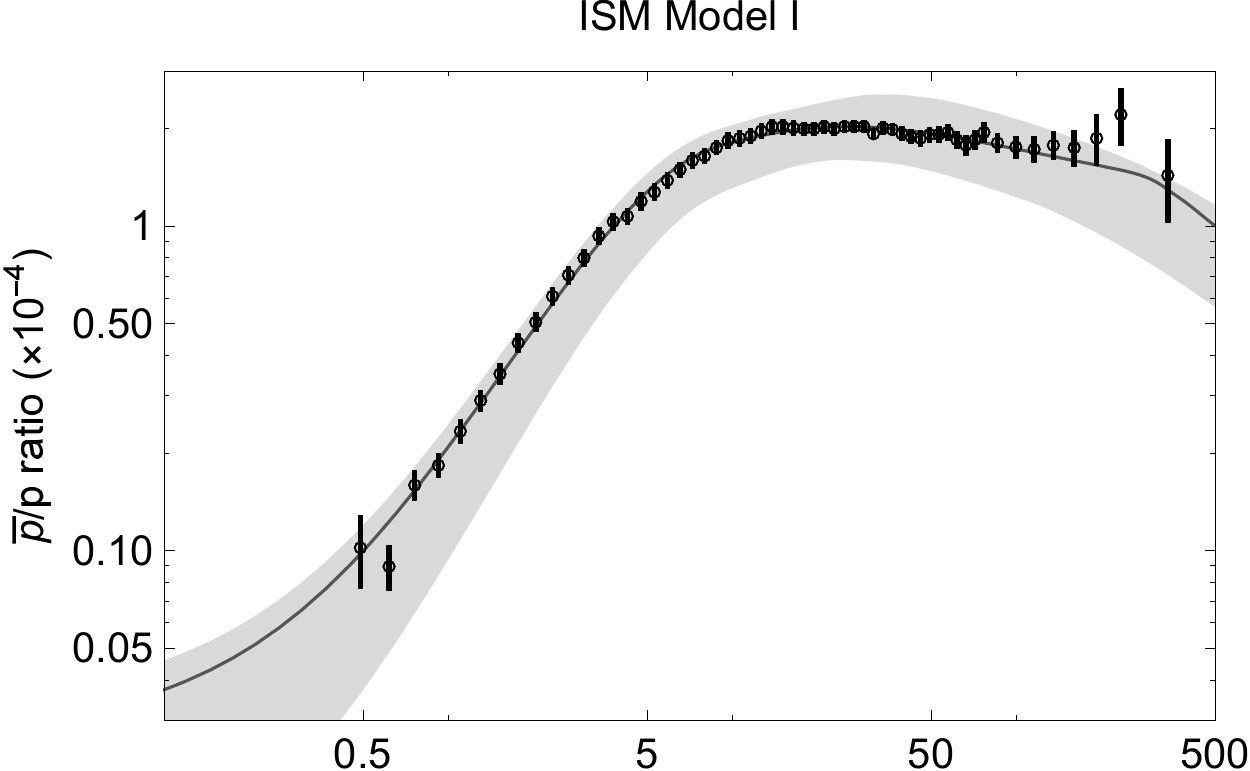}
\includegraphics[width=2.246in,angle=0]{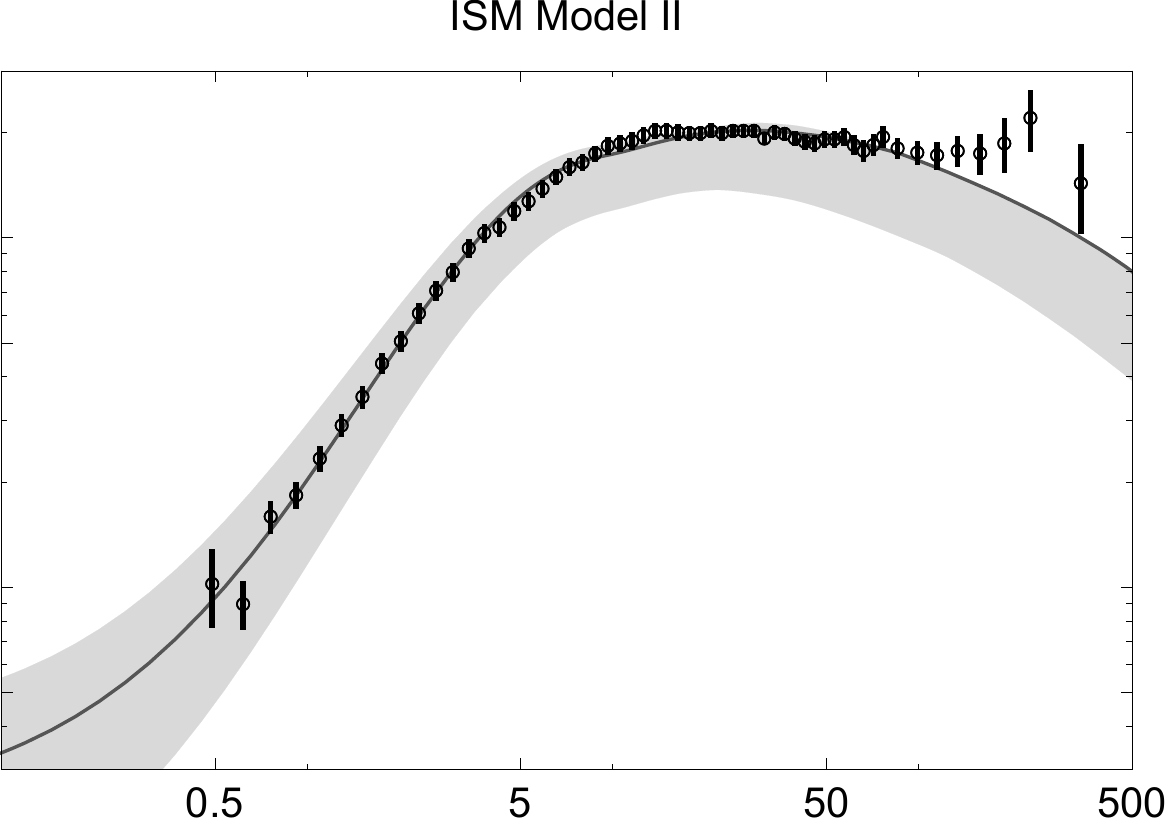}
\includegraphics[width=2.246in,angle=0]{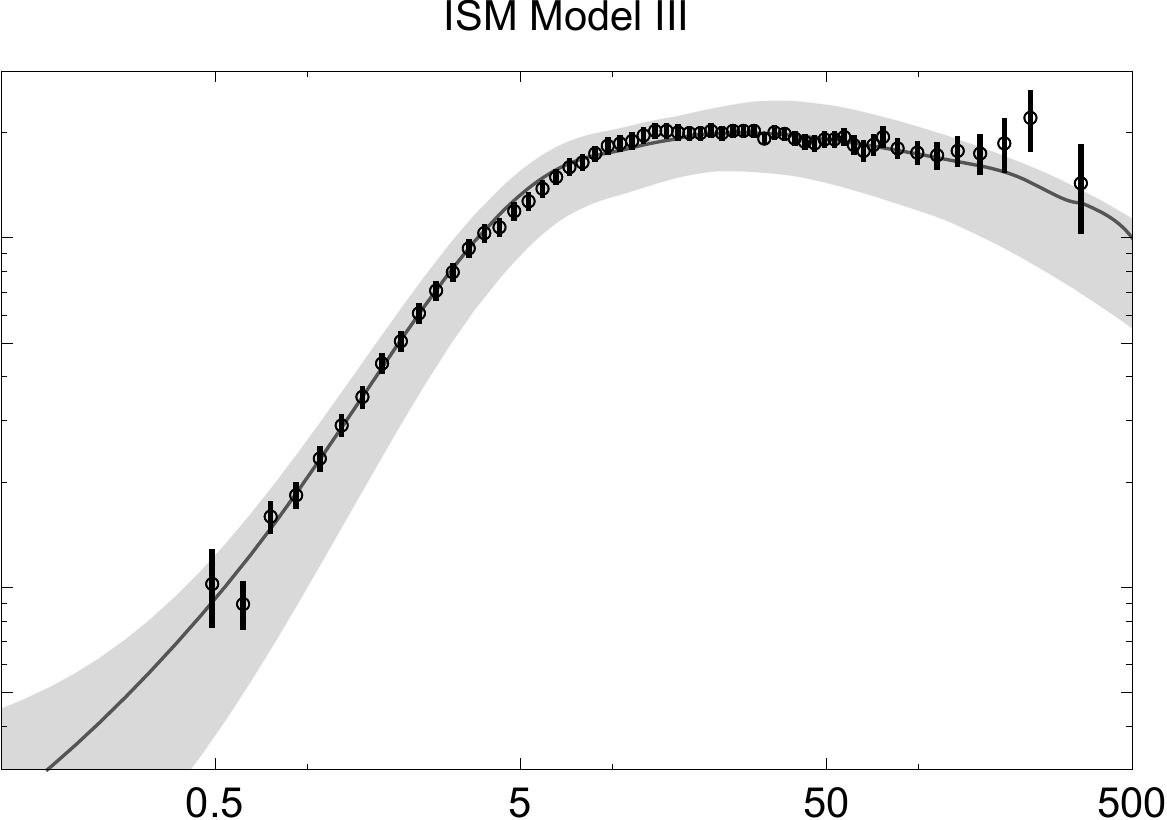}\\
\hspace{-0.0272in}
\hspace{-0.165in}
\includegraphics[width=2.505in,angle=0]{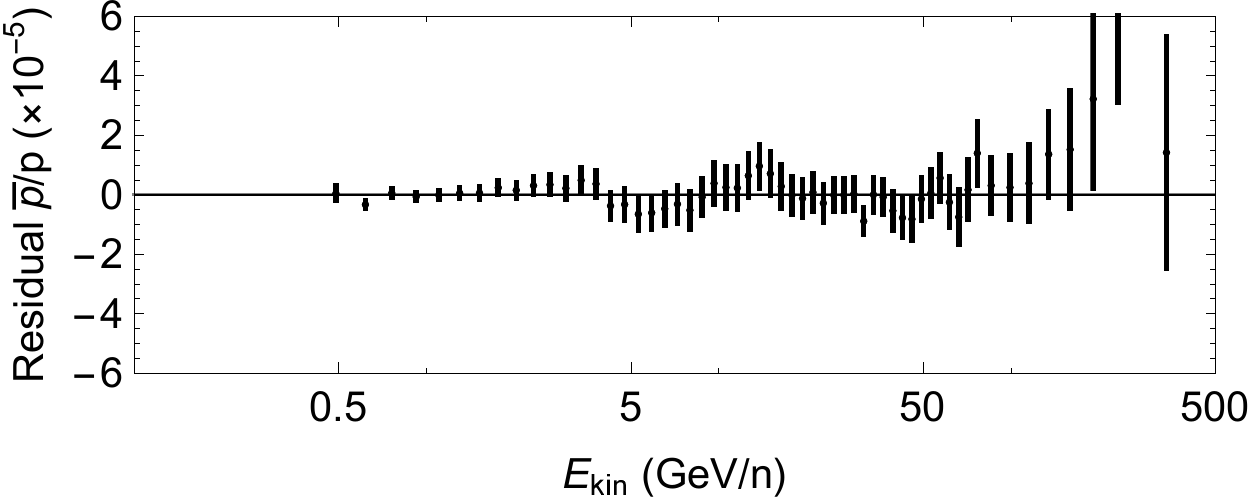}
\includegraphics[width=2.246in,angle=0]{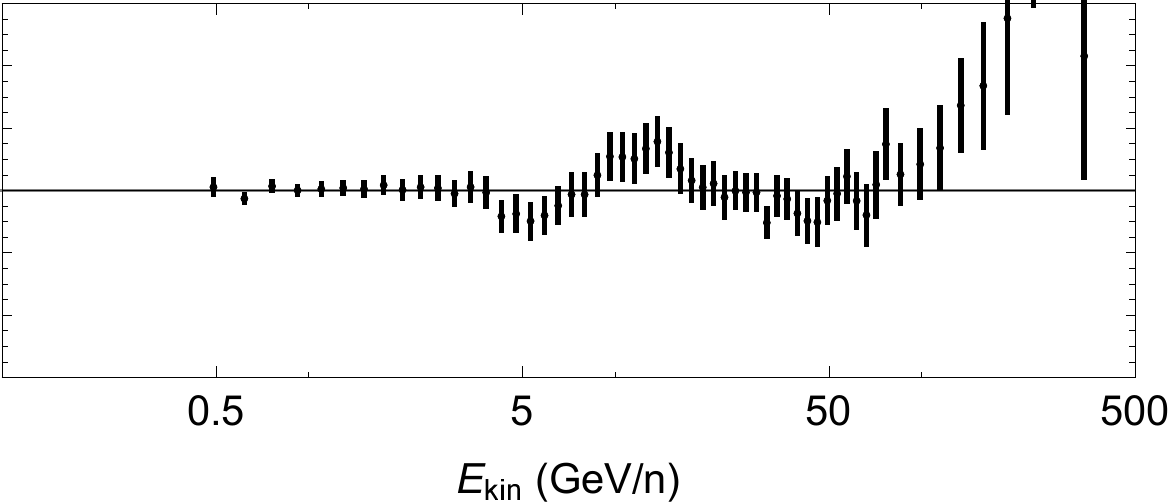}
\includegraphics[width=2.246in,angle=0]{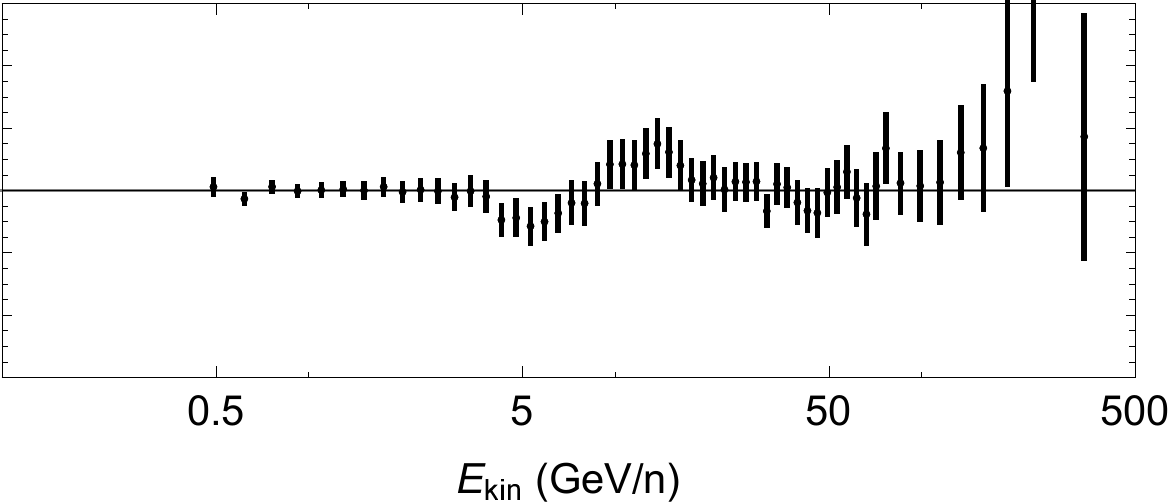}\\
\vskip -0.068in
\caption{The best-fit antiproton-to-proton ratio (gray solid line), without any contribution from annihilating dark matter. From left-to-right, each frame corresponds to a different model for the injection and propagation of cosmic rays in the ISM (see Table~\ref{tab:ISMBack}). The grey bands represent the combined uncertainties associated with solar modulation and the antiproton production cross section, which we marginalize over (and which are highly correlated between spectral bins). In the bottom panels, we show the difference between the measured and predicted values of the antiproton-to-proton ratio. The data points shown refer to the observations of \textit{AMS-02} as presented in Ref.~\cite{Aguilar:2016kjl}.}
\label{fig:pbar_NoDM_NoSAS}
\end{figure*}

\section{Results}
\label{sec:results}
\subsection{Fitting the Antiproton-to-Proton Ratio Without Dark Matter}

We begin by considering the antiproton-to-proton ratio presented by the \textit{AMS-02} Collaboration~\cite{Aguilar:2016kjl}, performing the fit without any  contribution from dark matter or other exotic physics. We treat the astrophysical and particle physics uncertainties as described in the previous section, and consider each model for the injection and propagation in the ISM independently (see Table~\ref{tab:ISMBack}). For each ISM model, we scan across a six-dimensional grid ($\phi_{0}$, $\phi_{1}$, $a$, $b$, $c$, and the normalization of the ISM gas density), calculating the log-likelihood for each point in this parameter space and then selecting the combination of values that provides the best fit to the data. The results of our fit are shown in Fig.~\ref{fig:pbar_NoDM_NoSAS}. The grey band shown in each frame represents the combined uncertainties associated with solar modulation and the antiproton production cross section. Although these bands are quite wide, we emphasize that these uncertainties are highly correlated, and are not generally capable of producing (or absorbing) narrow spectral features, such as those which might arise from annihilating dark matter. In each case, our model provides a reasonably good description of the data, yielding a $\chi^2$ per degree-of-freedom of 0.79, 1.28 and 1.03 for ISM Models I, II and III, respectively (as these error bars include both statistical and systematic uncertainties, we caution that one cannot use these values to formally address the quality of the fit). In the lower frames of this figure, however, one can easily identify a positive residual which appears at $\sim$10-20 GeV (and a deficit at $\sim$5-10 GeV), as well as an excess at energies above $\sim$100 GeV.

\begin{figure*}
\hspace{-0.2176in}
\includegraphics[width=2.2in,angle=0]{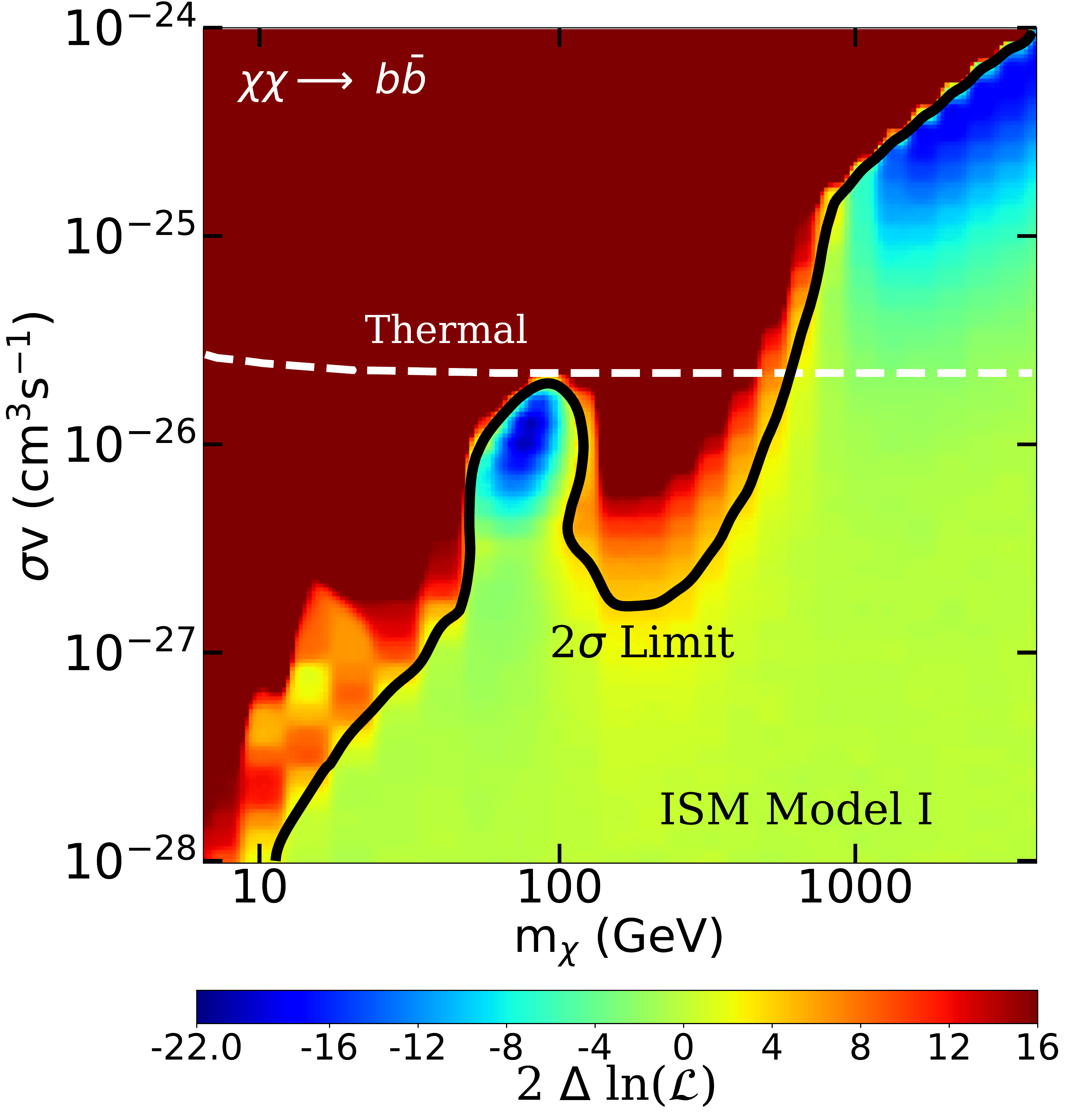} 
\hspace{0.1in}
\includegraphics[width=2.2in,angle=0]{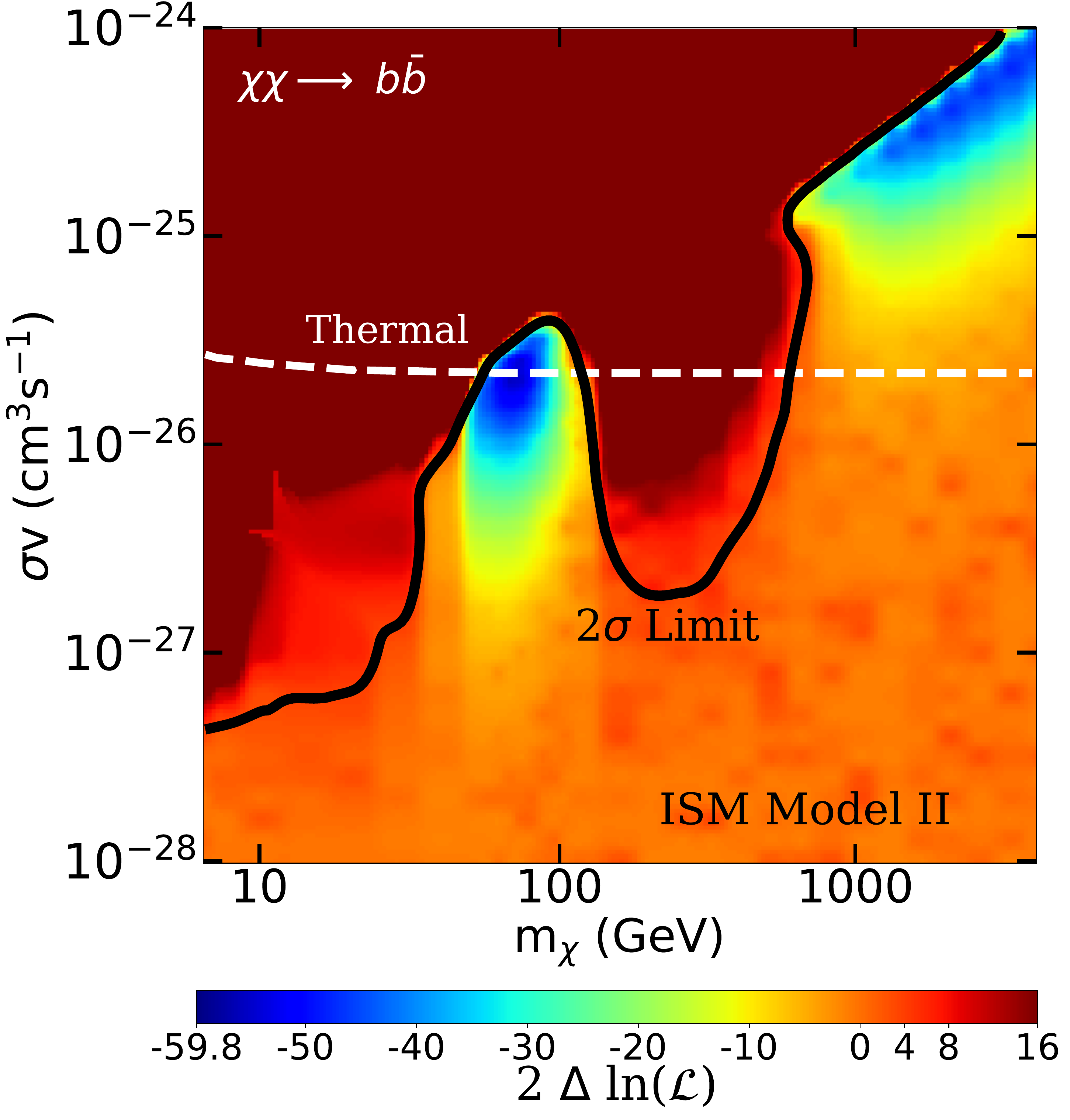}		
\hspace{0.1in}  
\includegraphics[width=2.2in,angle=0]{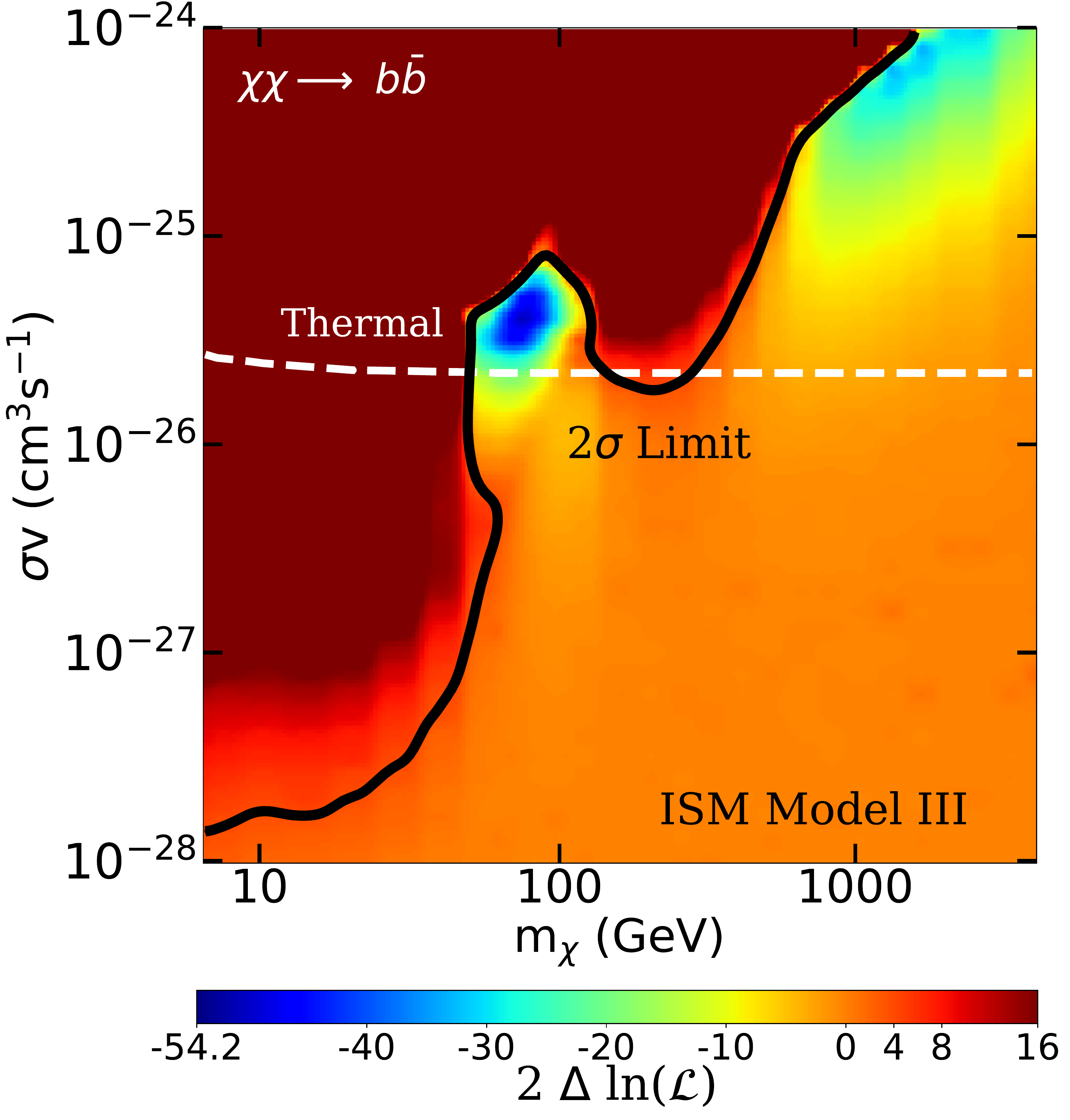}      
\vskip -0.068in
\caption{The impact of a contribution from annihilating dark matter on the log-likelihood of the fit to the \textit{AMS-02} antiproton-to-proton ratio, for the case of annihilations to $b\bar{b}$. Each frame corresponds to a different model for cosmic-ray injection and transport (see Table~\ref{tab:ISMBack}) and we have marginalized over the parameters associated with the antiproton production cross section and solar modulation (see Sec.~\ref{sec:method}). In each frame we find a statistically significant ($4.7\sigma$ or higher) preference for dark matter with $m_{\chi}=64-88$ GeV and $\sigma v = (0.7-5.2) \times 10^{-26}$ cm$^3$/s (see Table~\ref{tab:main}). The solid black curve represents the 2$\sigma$ upper limit on the annihilation cross section. The dashed white curve denotes the annihilation cross section predicted for dark matter in the form of a simple ($s-$wave) thermal relic. Note that the lowest value of $2\Delta \ln \mathcal{L}$ shown in the color bar represents the significance of the best-fit dark matter model in that frame.}
\label{fig:Tim_NoSAS}
\end{figure*}

\begin{figure*}
\hspace{-0.2176in}
\includegraphics[width=2.58in,angle=0]{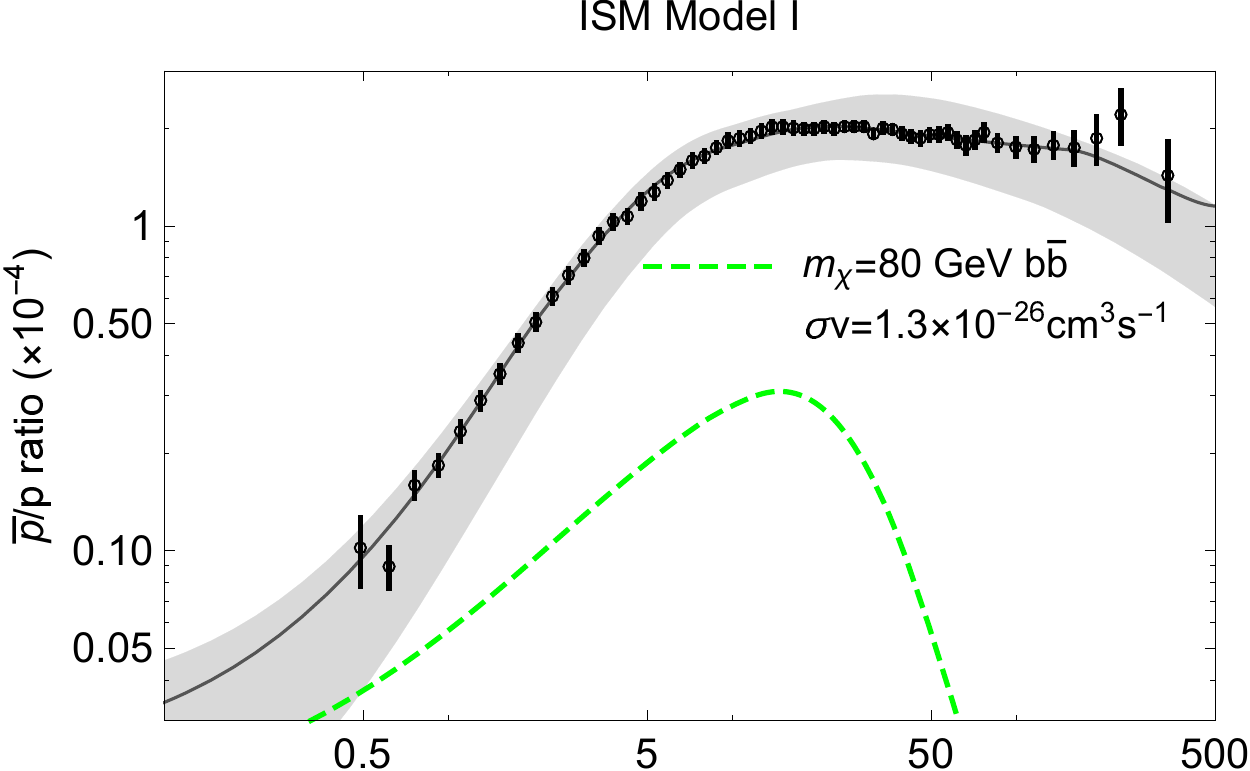}
\includegraphics[width=2.246in,angle=0]{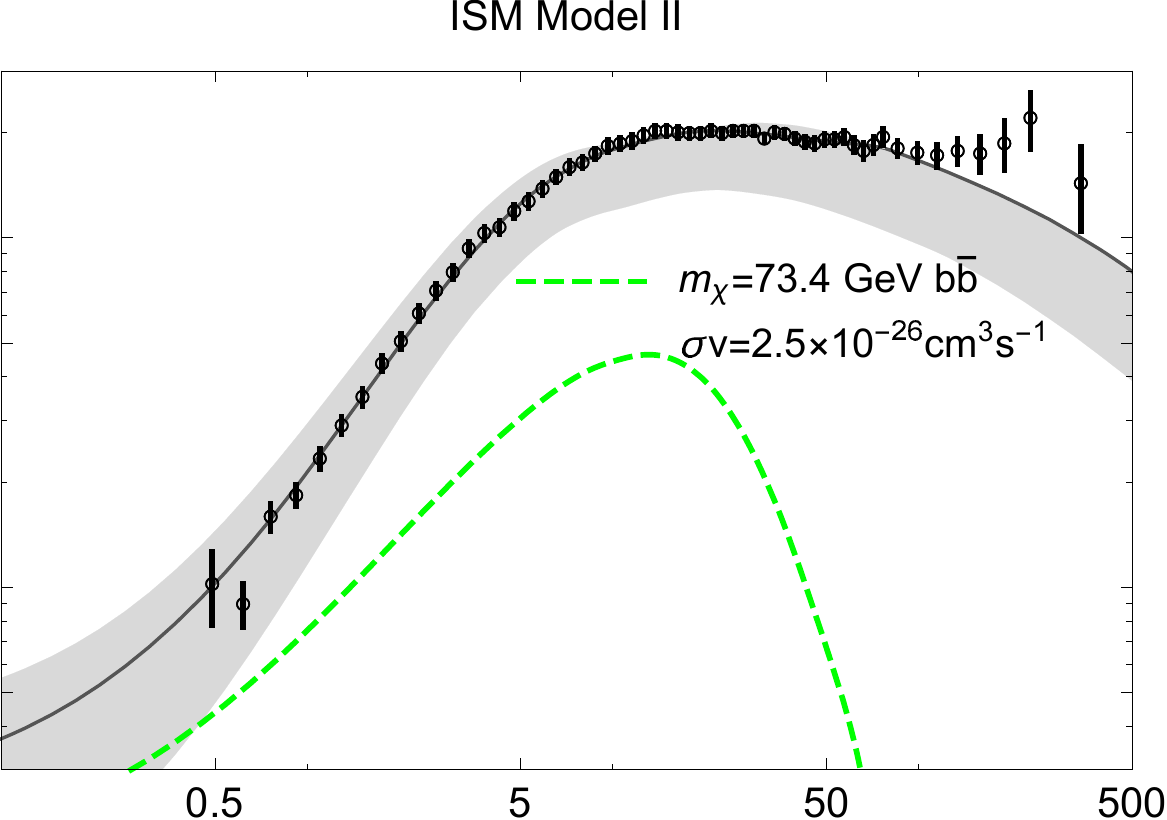}
\includegraphics[width=2.246in,angle=0]{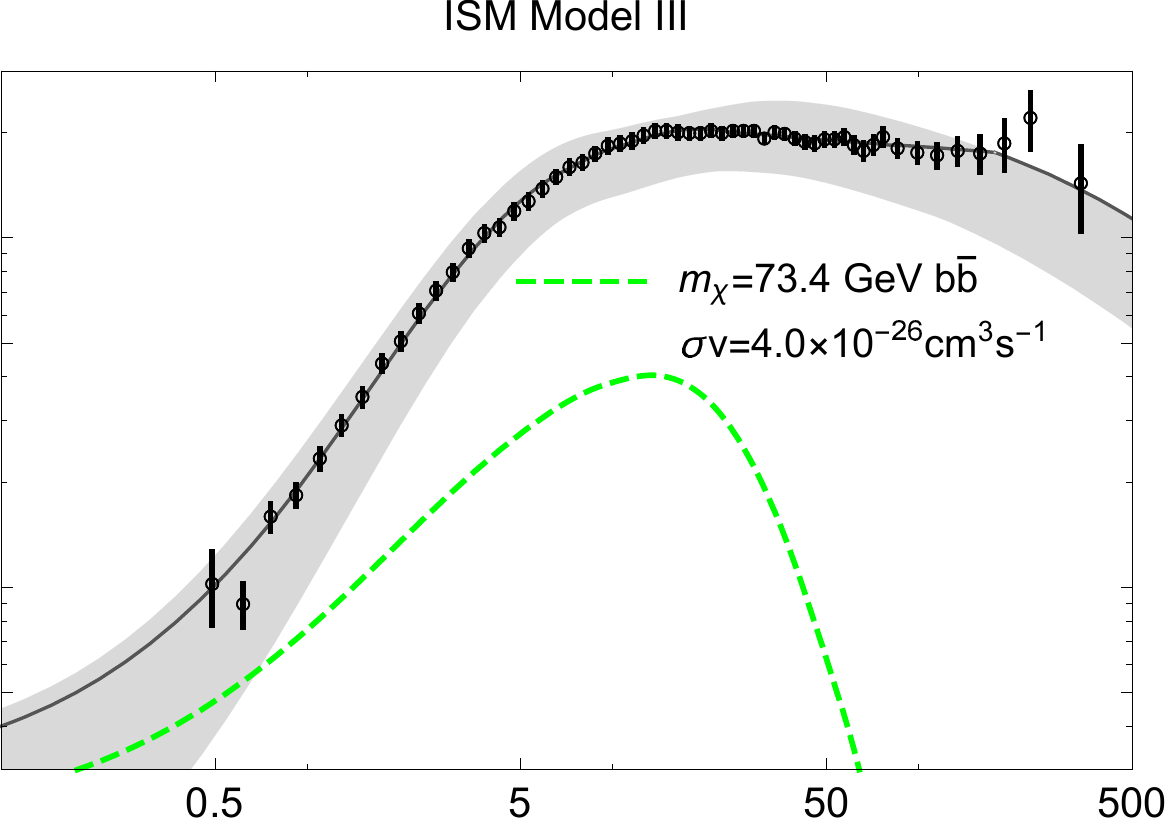}\\
\hspace{-0.0272in}
\hspace{-0.165in}
\includegraphics[width=2.505in,angle=0]{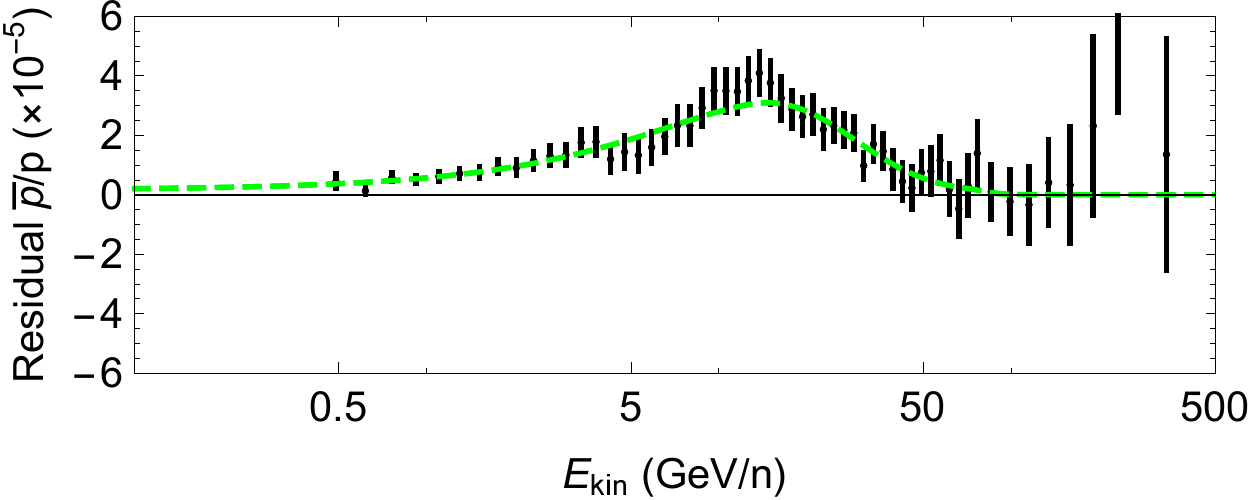}
\includegraphics[width=2.246in,angle=0]{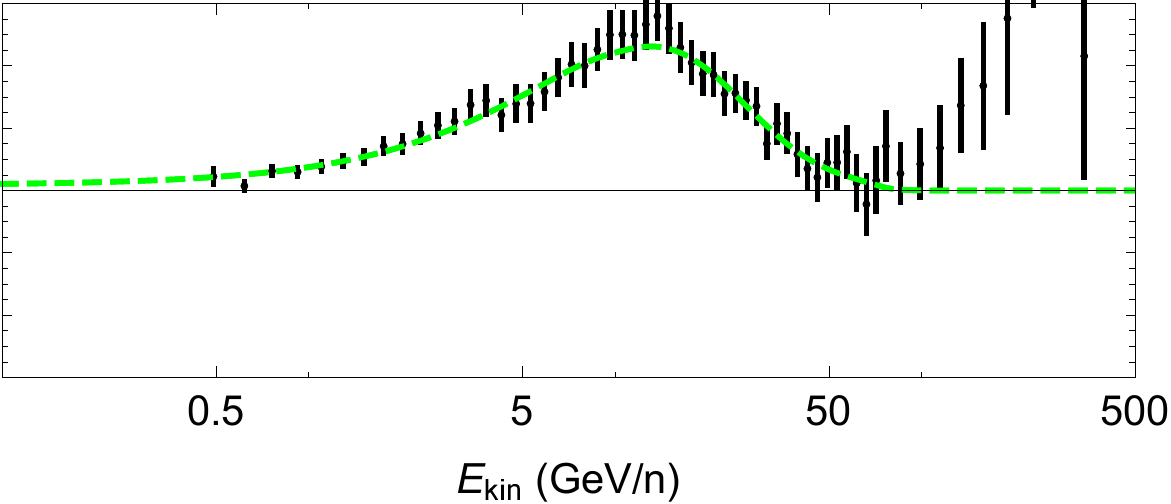}
\includegraphics[width=2.246in,angle=0]{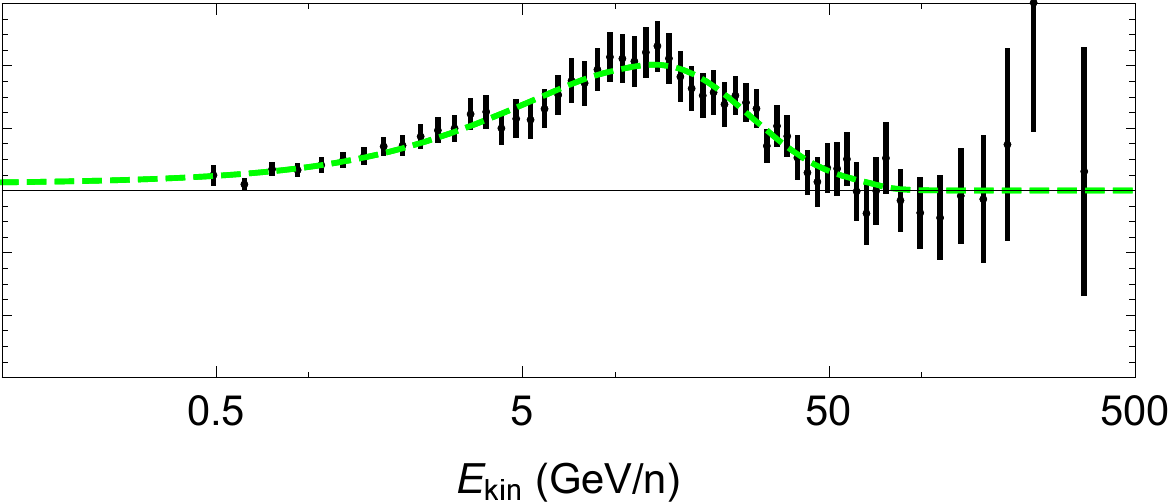}\\
\vskip -0.068in
\caption{As in Fig.~\ref{fig:pbar_NoDM_NoSAS}, but including the best-fit contribution from annihilating dark matter (shown in each frame as a green dashed line). In the lower frames, we plot the observed spectrum minus the astrophysical model, and thus these residuals include the best-fit contribution from annihilating dark matter.}
\label{fig:pbar_DM_NoSAS}
\end{figure*}   

\begin{figure*}
\hspace{-0.2176in}
\includegraphics[width=2.38in,angle=0]{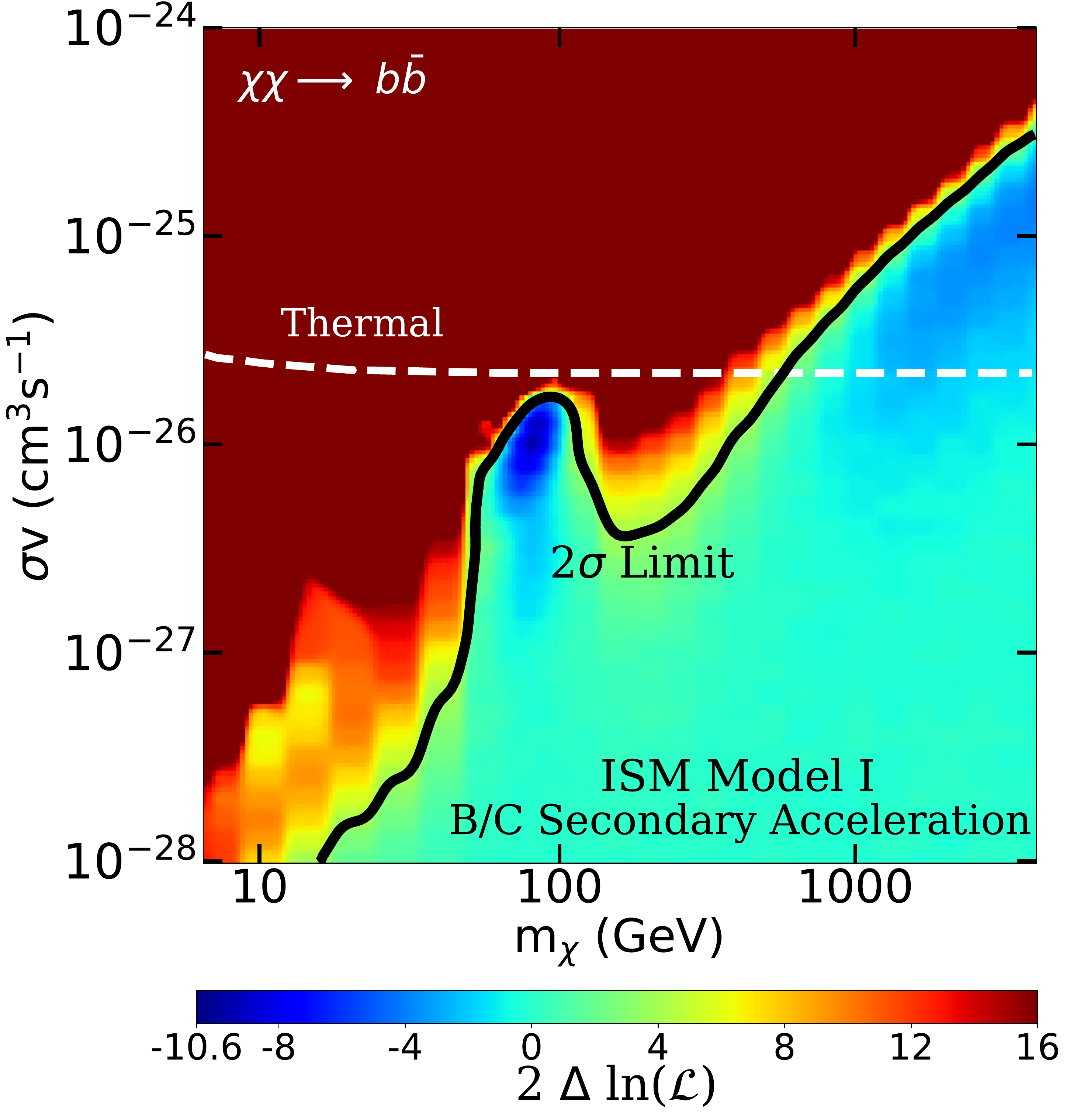} 
\includegraphics[width=2.38in,angle=0]{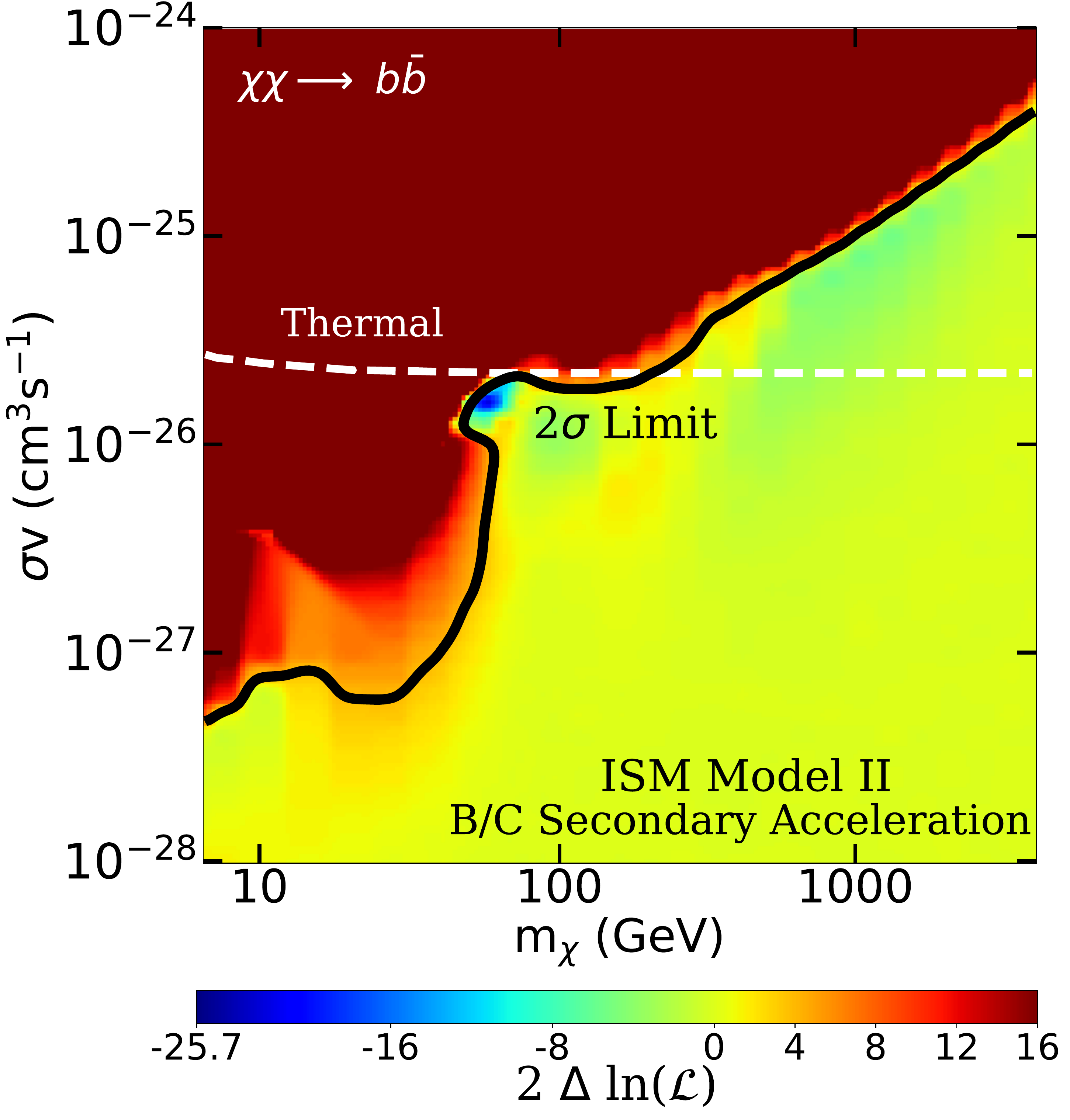}		
\includegraphics[width=2.38in,angle=0]{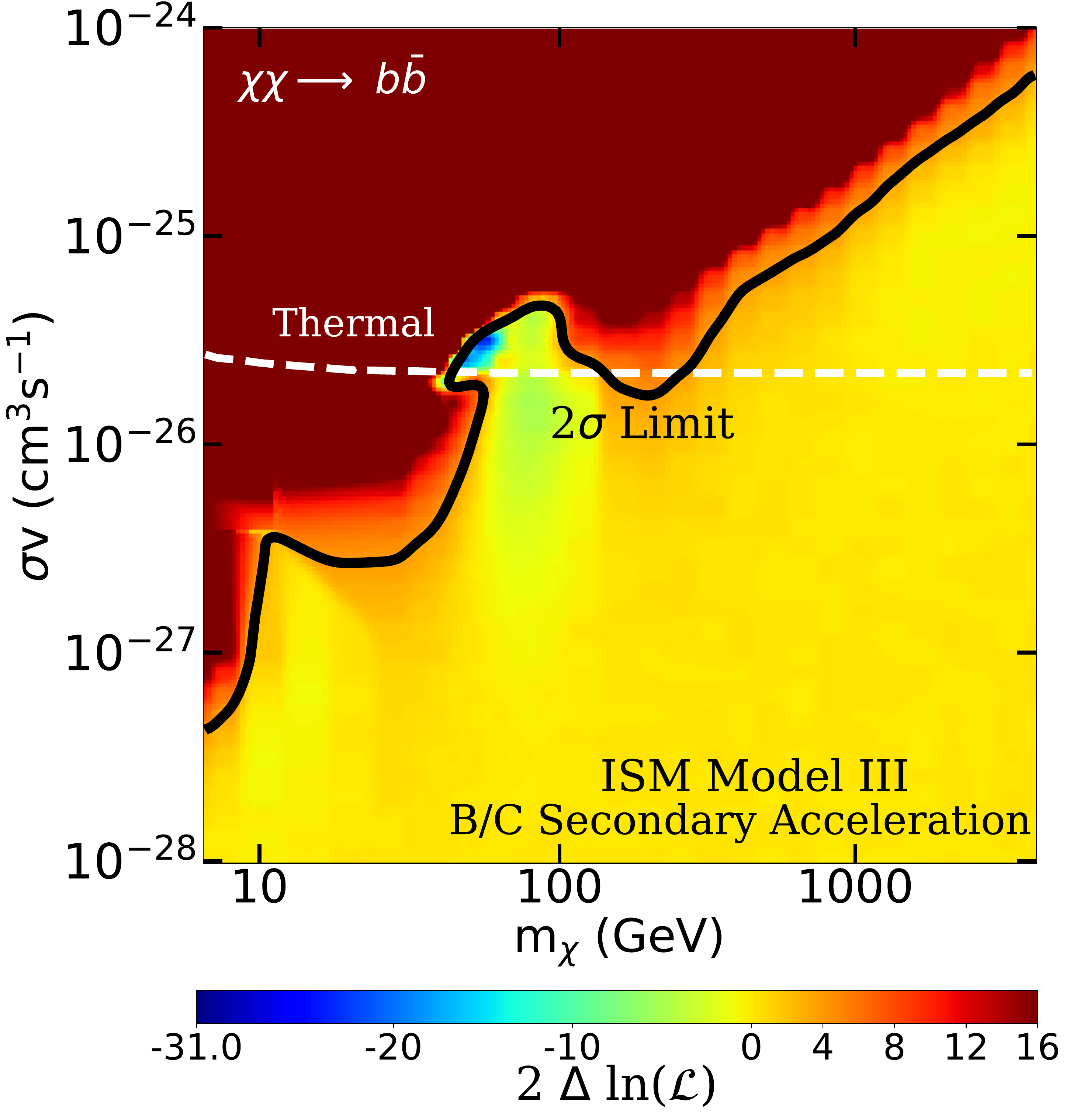}      
\vskip -0.068in
\caption{As in Fig.~\ref{fig:Tim_NoSAS}, but including a contribution from stochastically accelerated secondary antiprotons with values of $K_B$ and $n_{\rm gas}$ selected in order to provide a good fit the measured boron-to-carbon ratio. The presence of the accelerated secondaries largely removes the excess above $\sim$100 GeV, erasing the preference seen in Fig.~\ref{fig:Tim_NoSAS} for a $\sim$1-3 TeV dark matter particle. The preference for a lighter dark matter particle largely persists, favoring $m_{\chi}=46-94$ GeV and $\sigma v = (0.7-3.3) \times 10^{-26}$ cm$^3$/s with a statistical significance of 3.3$\sigma$ (see Table~\ref{tab:main}).}
\label{fig:Tim_WithSASBC}
\end{figure*}

\begin{figure*}
\hspace{-0.2176in}
\includegraphics[width=2.58in,angle=0]{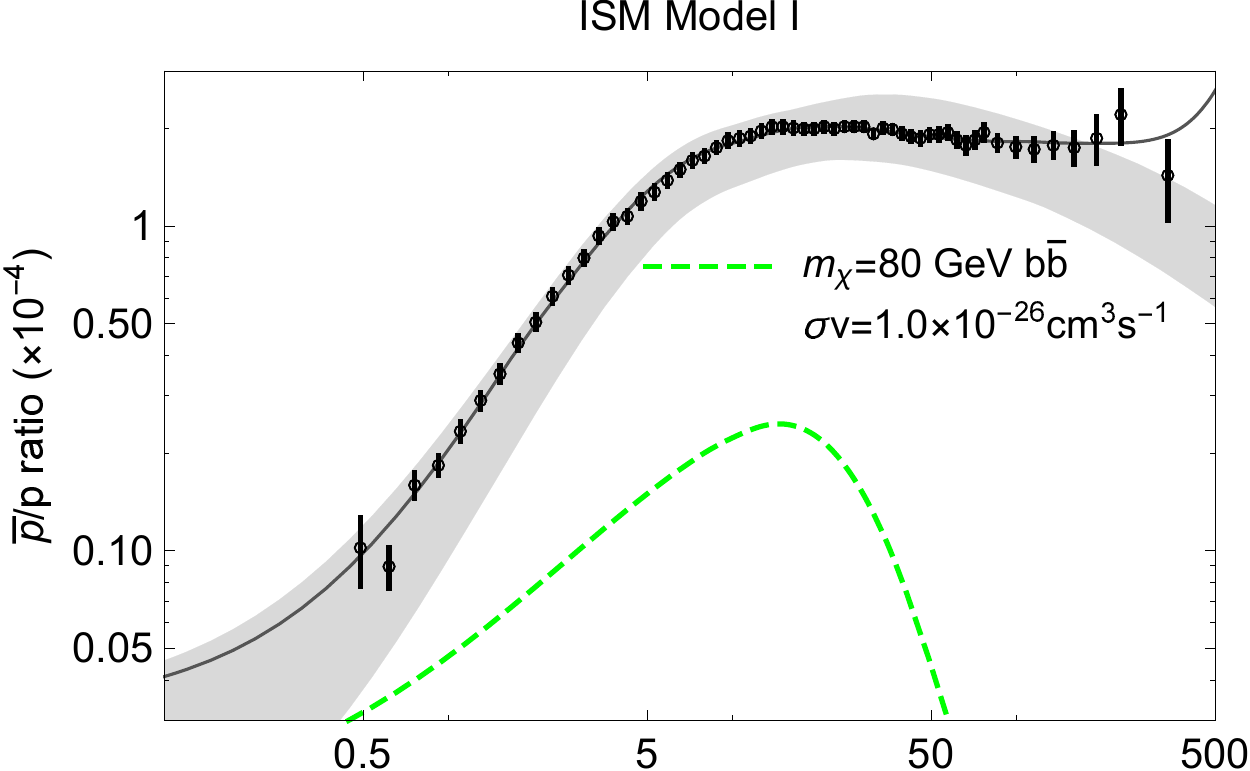}
\includegraphics[width=2.246in,angle=0]{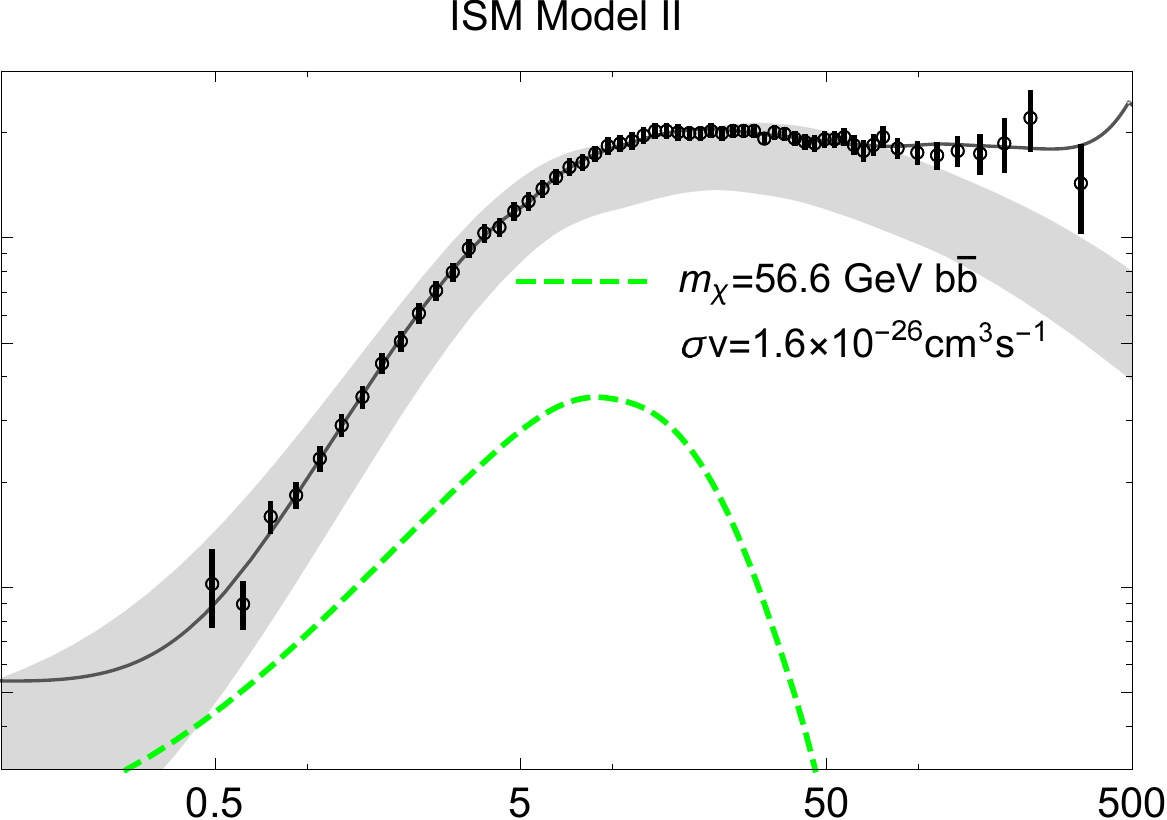}
\includegraphics[width=2.246in,angle=0]{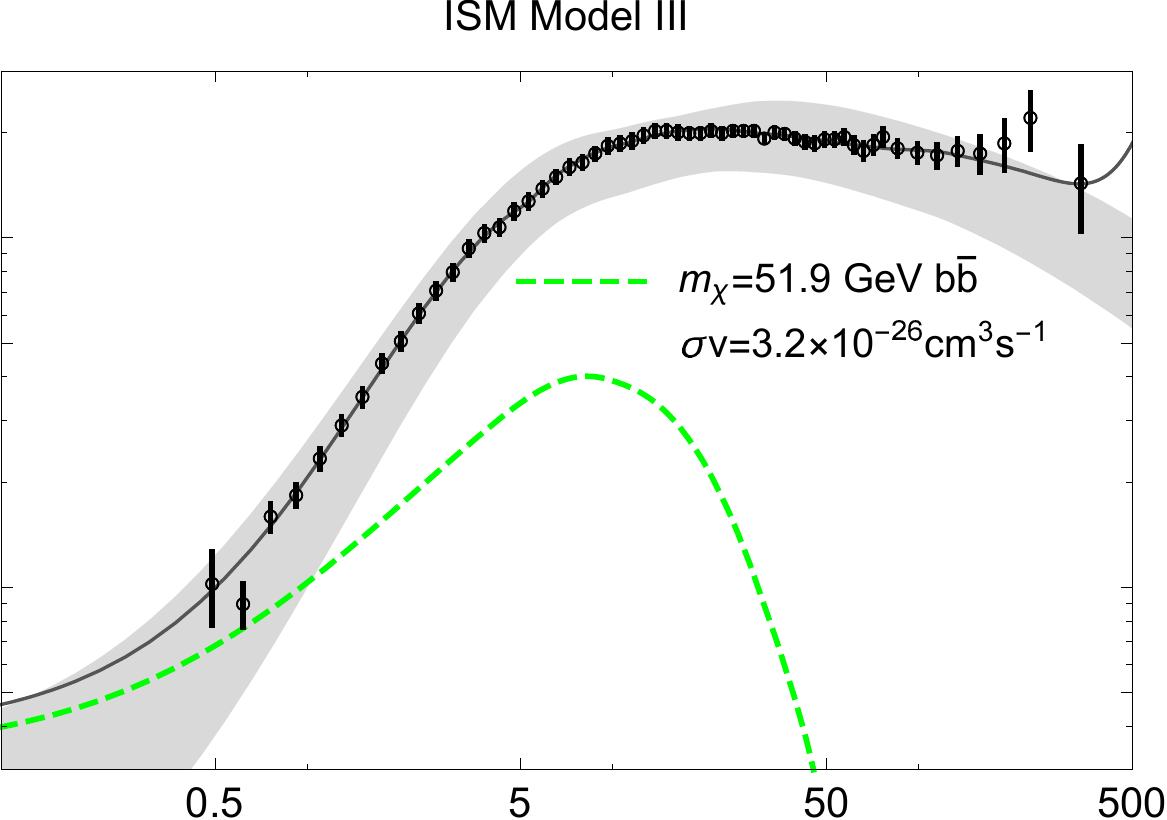}\\
\hspace{-0.0272in}
\hspace{-0.165in}
\includegraphics[width=2.505in,angle=0]{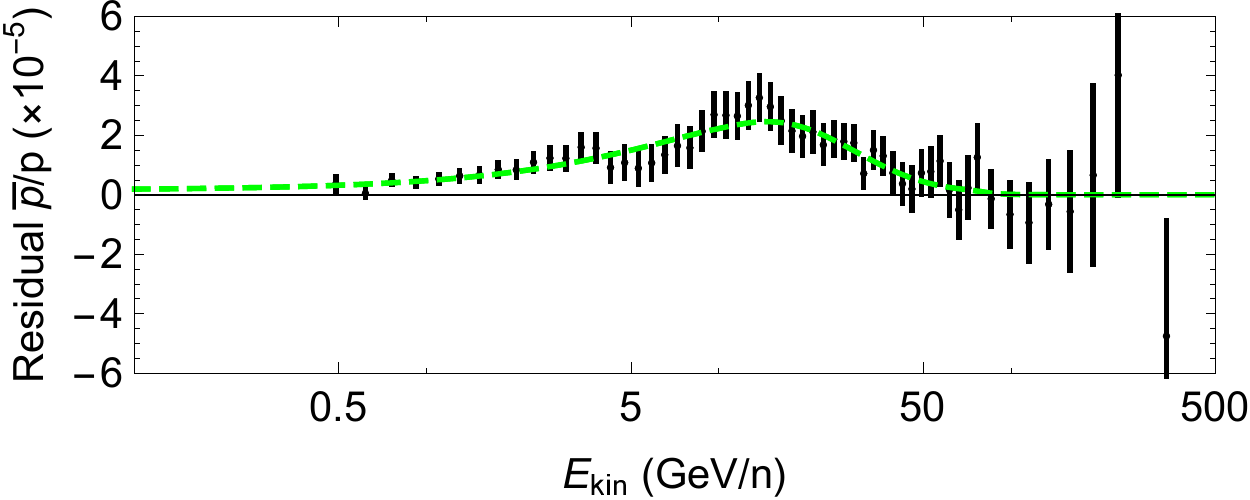}
\includegraphics[width=2.246in,angle=0]{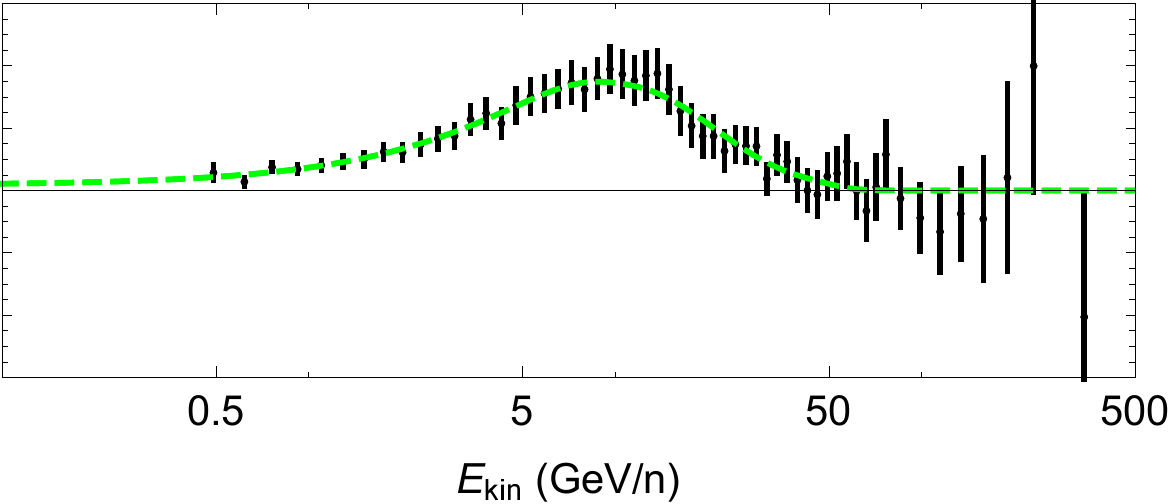}
\includegraphics[width=2.246in,angle=0]{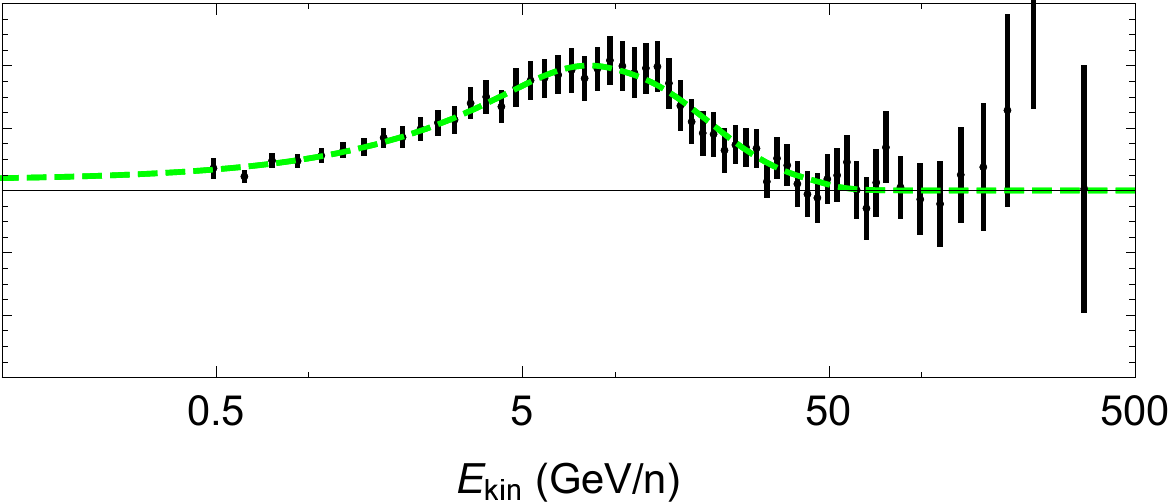}\\
\vskip -0.068in
\caption{As in Fig.~\ref{fig:pbar_DM_NoSAS}, but including a contribution from stochastically accelerated secondary antiprotons with a values of $K_B$ and $n_{\rm gas}$ selected in order to provide a good fit the measured boron-to-carbon ratio.}
\label{fig:pbar_DM_WithSASBC}
\end{figure*}

\subsection{Including Annihilating Dark Matter}
\label{sec:DMA} 

The spectrum of antiprotons produced in dark matter annihilation processes can be calculated using Monte Carlo event generators such as \texttt{PYTHIA}~\cite{Sjostrand:2007gs} and \texttt{HERWIG}~\cite{Corcella:2000bw}. In this study, we use the publicly 
available \texttt{PPPC4DMID} code~\cite{Cirelli:2010xx} which provides the differential 
spectra of antiprotons from DM annihilations, $dN_{\bar{p}}/dE_{\bar{p}}$. Although throughout most of this study we focus on the representative case of annihilations to $b\bar{b}$, we consider in the Appendix models in which the dark matter annihilates to light quarks or to $W^+ W^-$. The \texttt{PPPC4DMID} code includes electroweak corrections which are important in the case of heavy dark matter particles, when the annihilation products can be highly boosted and emit a $W$ or $Z$ before decaying or hadronizing~\cite{Ciafaloni:2010ti}.

For the distribution of dark matter in the Milky Way we adopt an Navarro-Frenk-White 
(NFW) profile~\cite{Navarro:1995iw}:
\begin{equation}
    \rho(r) = \frac{\rho_0}{(r/r_s)(1 + r/r_s)^{2}}.
    \label{eq:NFW}
\end{equation}
We set the normalization parameter, $\rho_{0}$, such that the local density (at $r=8.5$ kpc) is 0.4 GeV/cm$^3$~\cite{Catena:2009mf, 
Salucci:2010qr} and adopt a scale radius of $r_s = 20$ kpc. We note that the results presented here are not highly sensitive to the choice of the halo profile. If we had instead adopted an Einasto profile~\cite{Einasto} or a profile with a slightly steeper inner slope (as motivated by the observed profile of the Galactic Center gamma-ray excess~\cite{Daylan:2014rsa,Calore:2014xka}), the local antiproton spectrum would be largely unaffected. The reason for this is that most of the cosmic rays in the energy range of interest originate from the surrounding few kpc, and thus the dependence on the dark matter halo profile is largely limited to the overall normalization (\ie the local density). 

In Fig.~\ref{fig:Tim_NoSAS}, we show the impact of annihilating dark matter on the fit to the antiproton-to-proton ratio for the case of dark matter annihilating to $b\bar{b}$. For the case of the thermal relic benchmark cross section (shown as a white dashed line~\cite{Steigman:2012nb}), this data excludes (at the $2\sigma$ level) dark matter masses up to 47 GeV and between 136-286 GeV, representing one of the strongest constraints on annihilating dark matter. There are two regions of parameter space, however, in which a dark matter annihilation signal improves the quality of the fit. The best overall fit is found for the case in which a $m_{\chi}=64-88$ GeV dark matter candidate annihilates with a cross section of $\sigma v = (0.8-5.2) \times 10^{-26}$ cm$^3$/s. Such a contribution improves the fit by $2 \Delta \ln \mathcal{L} = 22.0$, 59.8 and 54.2 for ISM Models I, II, and III, respectively, corresponding to a statistical preference between 4.7 and 7.7$\sigma$. It is noteworthy how similar these parameters are to those that are required to generate the observed characteristics of the Galactic Center gamma-ray excess~\cite{Daylan:2014rsa,Calore:2014xka}. At higher masses ($\gsim 1$ TeV), annihilating dark matter particles can also improve the fit to this dataset, although to a lesser extent. We remind the reader that at each point in the fit we have marginalized over the parameters associated with the antiproton production cross section and solar modulation as described in Sec.~\ref{sec:method}, and therefore our results indicate that the presence of this excess is statistically significant, even in light of these systematic uncertainties.

In Fig.~\ref{fig:pbar_DM_NoSAS}, we show the spectrum of the antiproton-to-proton ratio, including the best-fit contribution from annihilating dark matter. The residual plots (lower frames) clearly illustrate the preference for a contribution from annihilating dark matter peaking in at energies near $\sim$10-20 GeV. In the top three rows of Table~\ref{tab:main} we summarize our results, listing the values of the dark matter mass and annihilation cross section that are favored by this fit, for each of the three cosmic-ray injection and transport models considered in this study. In each case, we find a statistically significant preference for a contribution from annihilating dark matter.

We note that our analysis arrives at qualitatively different conclusions than those presented in Ref.~\cite{Reinert:2017aga}, which finds that the statistical significance of the antiproton excess can be reduced to approximately 2.2$\sigma$ after systematic uncertainties are taken into account. We note that there are many significant differences between the cosmic-ray propagation models employed between these papers. Most notably, the authors of Ref.~\cite{Reinert:2017aga} utilize an analytic two-zone cosmic-ray propagation model, with parameters that are tuned to the antiproton data, as well as to the cosmic-ray positron flux. We utilize numerical cosmic-ray propagation models based on the \texttt{Galprop} code, and choose not to normalize our astrophysical background models to cosmic-ray antiprotons (to avoid biasing the results) or to the spectrum of cosmic-ray leptons (which have vastly different cooling times). In addition, there are significant differences in our modeling of the antiproton production cross section and in our treatment of solar modulation. 

Up to this point, we have not considered the possibility that secondary antiprotons could be accelerated in the environments surrounding supernova remnants~\cite{Cholis:2017qlb,Blasi:2009hv, Mertsch:2009ph, Cholis:2013lwa, Kohri:2015mga, Fujita:2009wk}. In the following section we will consider how such a contribution could impact our results.


\begin{figure*}
\hspace{-0.2176in}
\includegraphics[width=2.38in,angle=0]{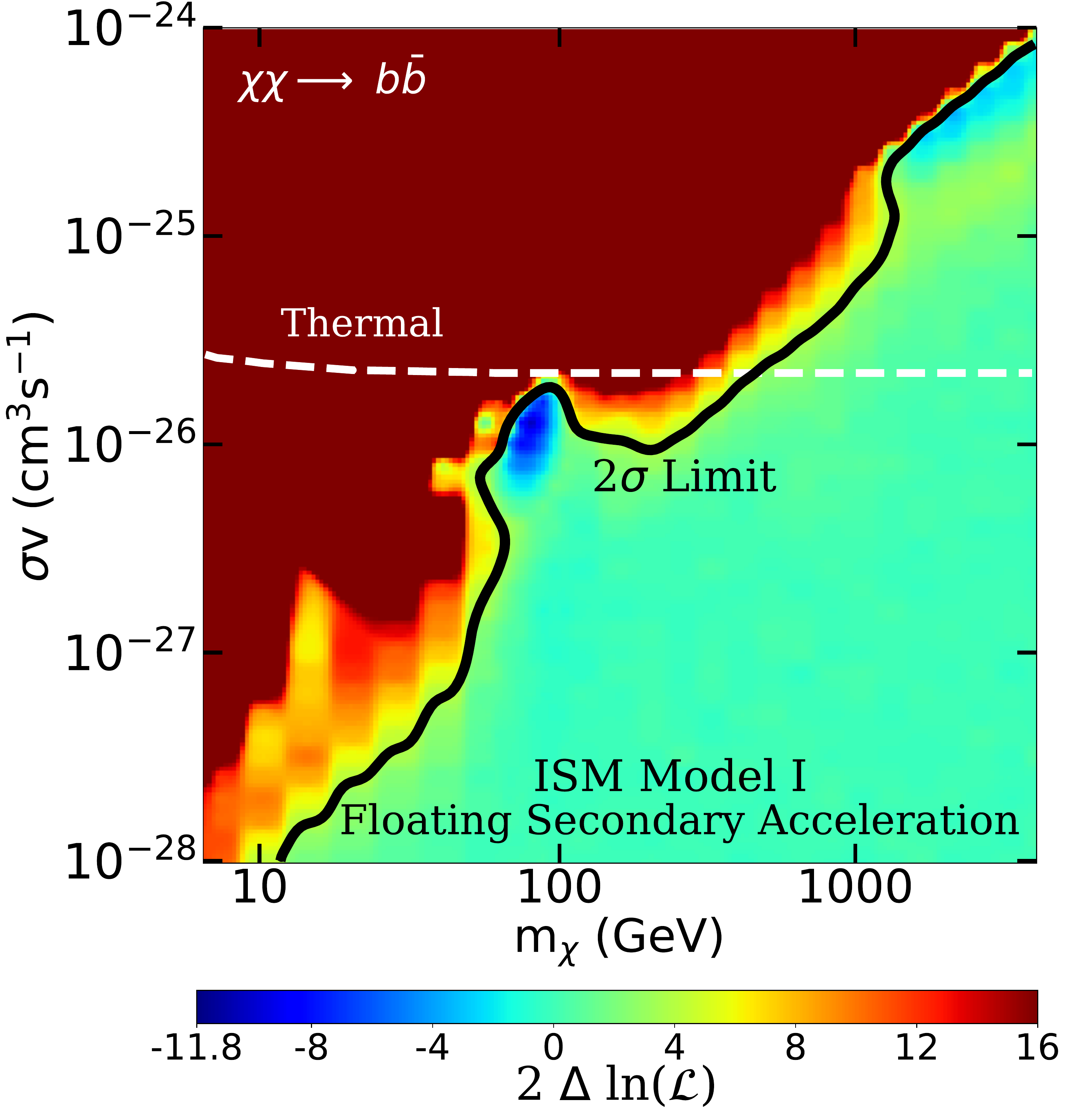} 
\includegraphics[width=2.38in,angle=0]{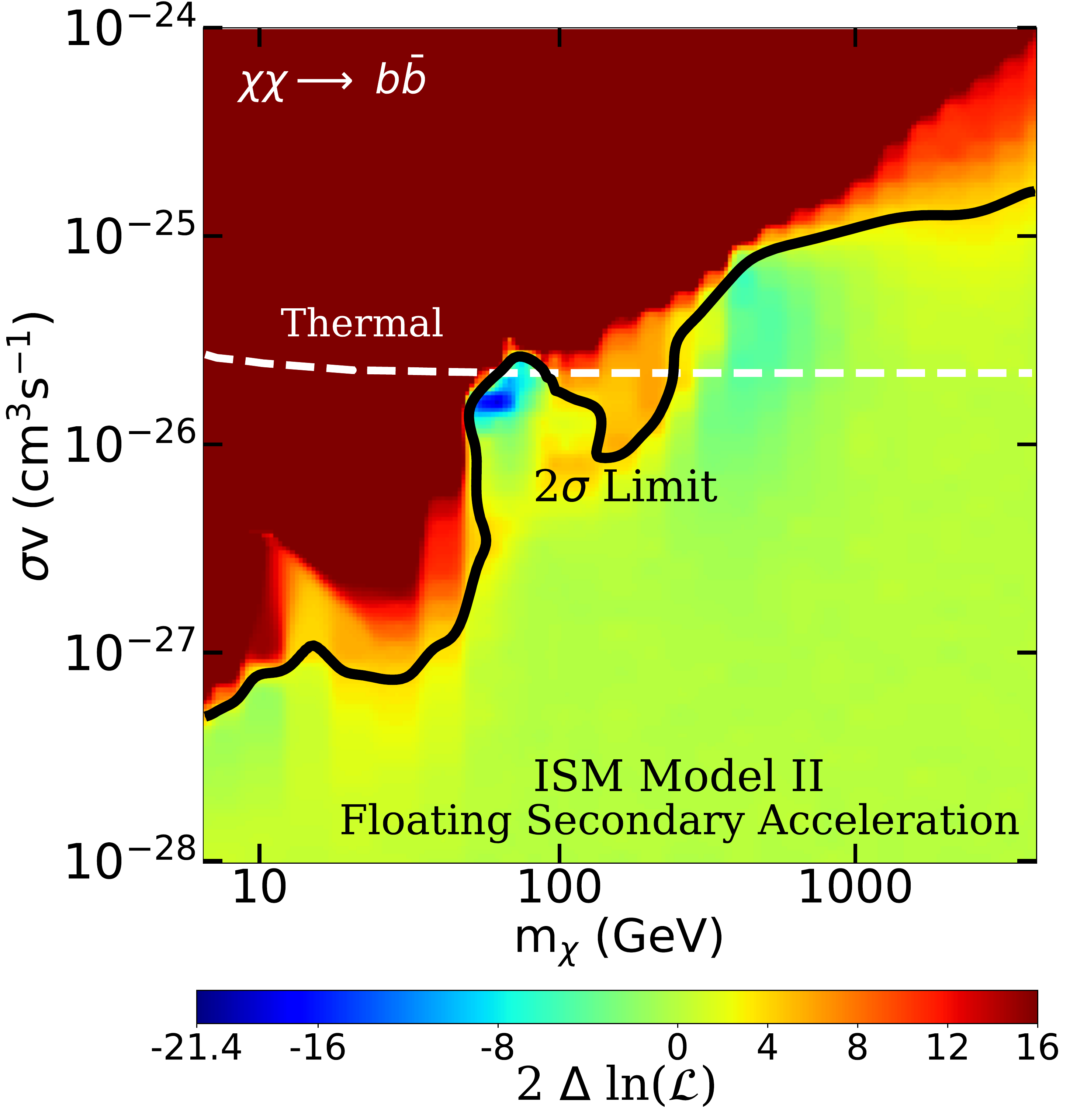}		
\includegraphics[width=2.38in,angle=0]{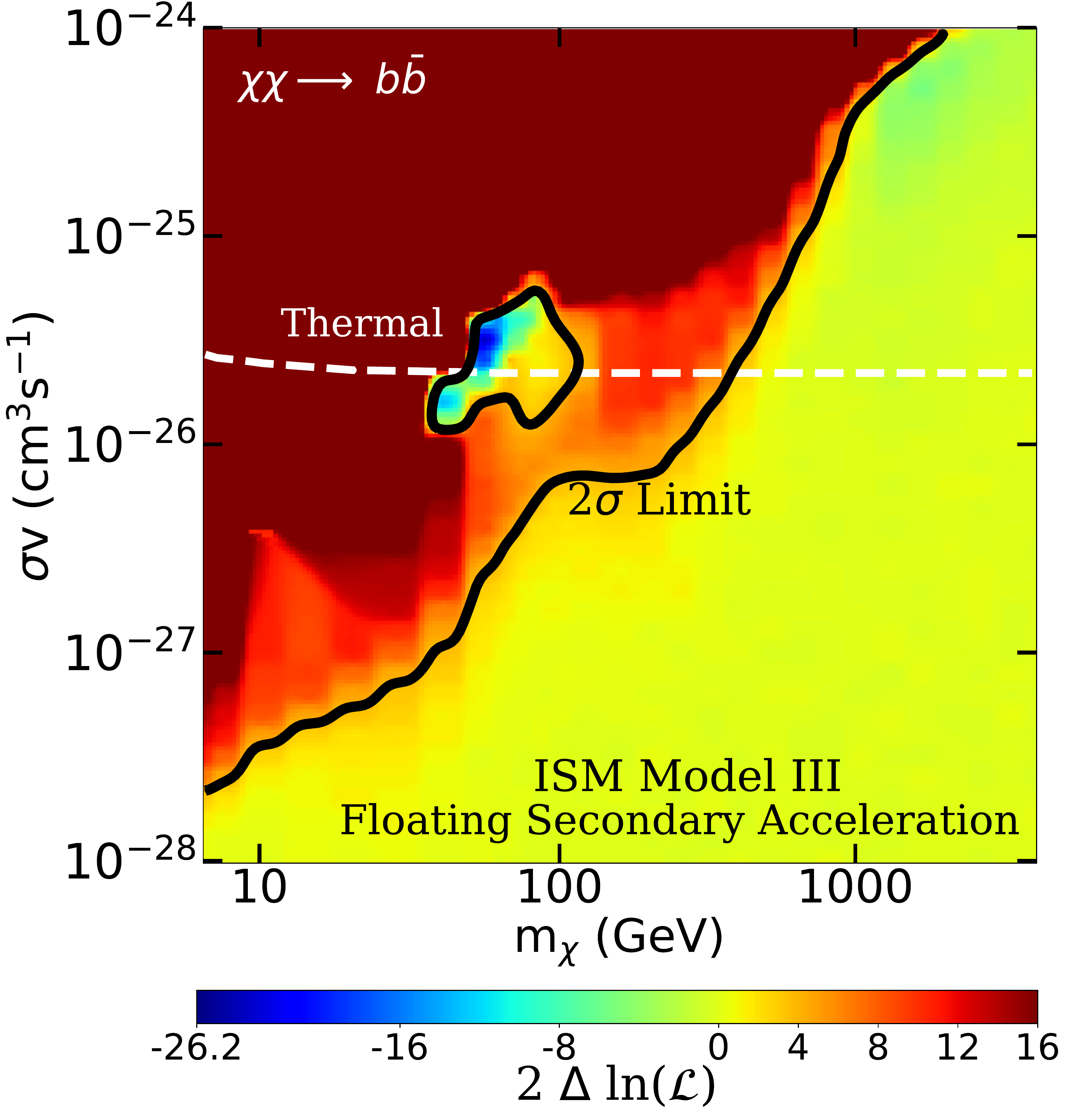}      
\vskip -0.068in
\caption{As in Figs.~\ref{fig:Tim_NoSAS} and~\ref{fig:Tim_WithSASBC}, but including a contribution from stochastically accelerated secondary antiprotons with a freely floating value of $n_{\rm gas}$. The presence of the accelerated secondaries largely removes the excess above $\sim$100 GeV, erasing the preference seen in Fig.~\ref{fig:Tim_NoSAS} for a $\sim$1-3 TeV dark matter particle. The preference for a lighter dark matter particle largely persists, favoring $m_{\chi}=46-89$ GeV and $\sigma v = (0.9-3.8) \times 10^{-26}$ cm$^3$/s with a statistical significance of 3.4$\sigma$ (see Table~\ref{tab:main}).}
\label{fig:Tim_WithSAS}
\end{figure*}

\begin{figure*}
\hspace{-0.2176in}
\includegraphics[width=2.58in,angle=0]{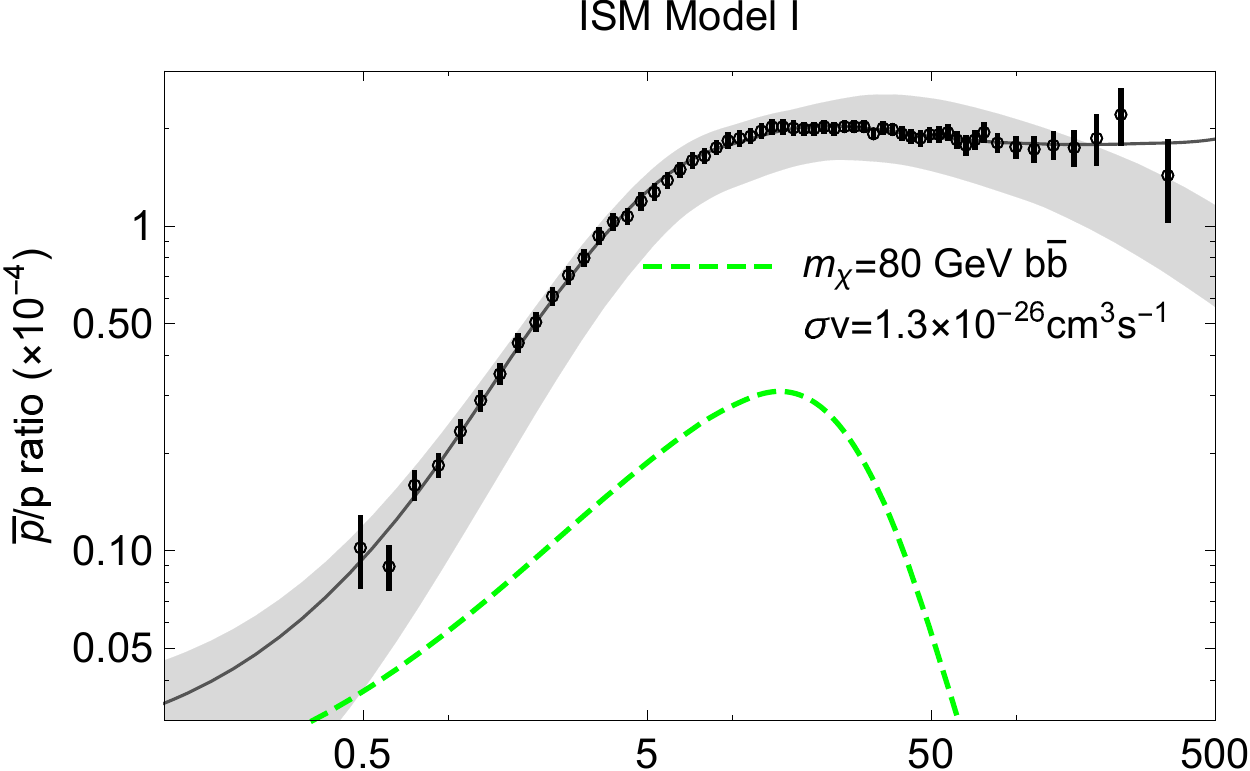}
\includegraphics[width=2.246in,angle=0]{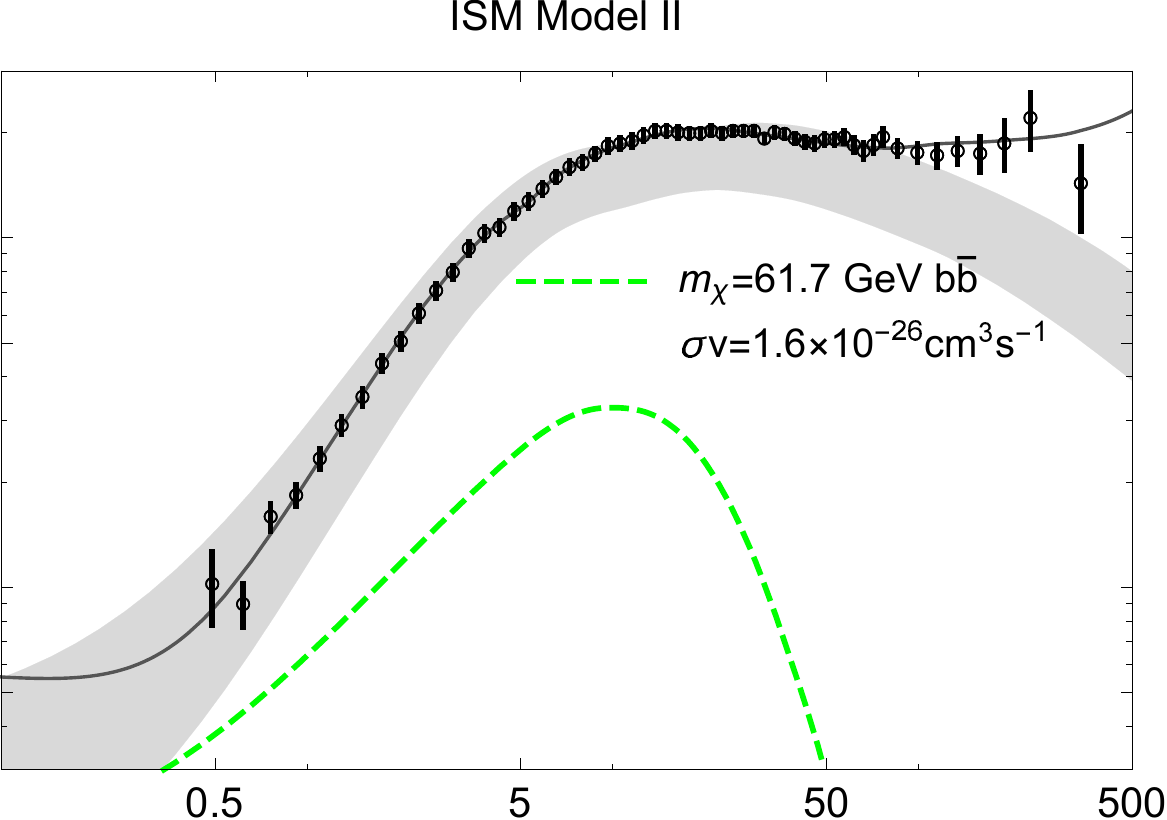}
\includegraphics[width=2.246in,angle=0]{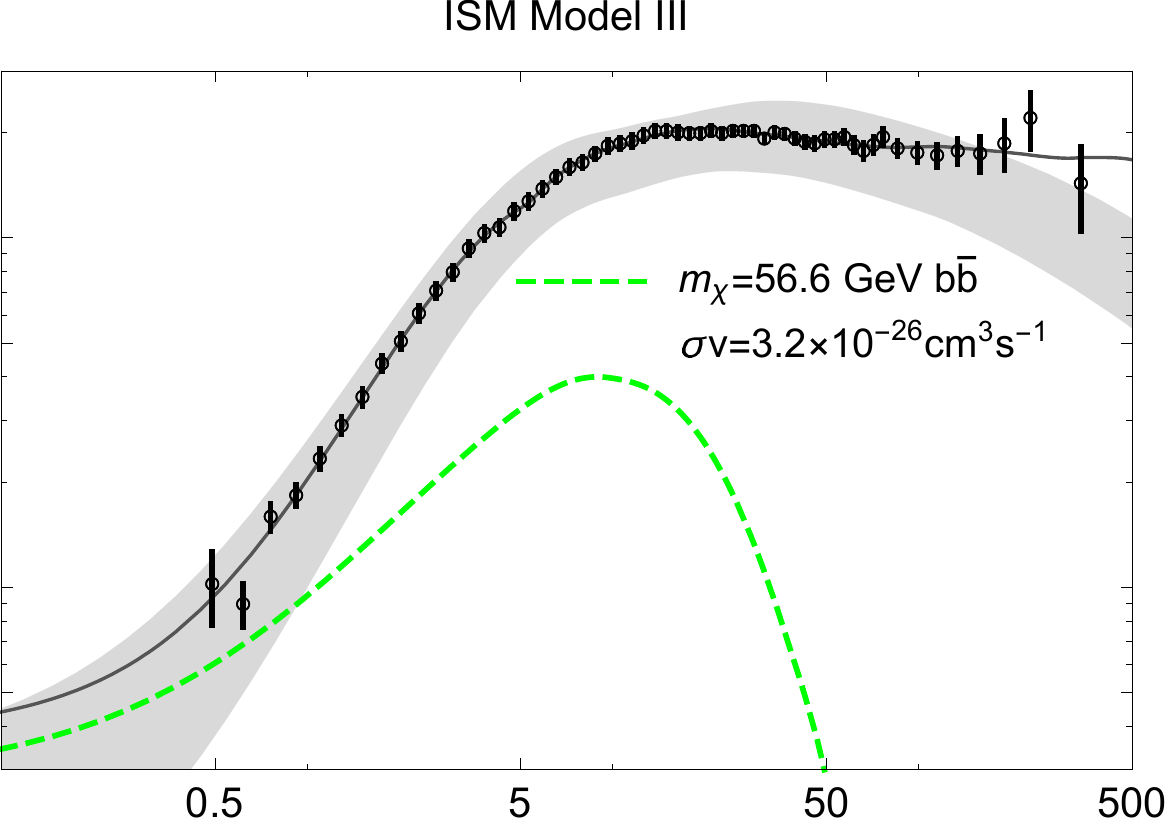}\\
\hspace{-0.0272in}
\hspace{-0.165in}
\includegraphics[width=2.505in,angle=0]{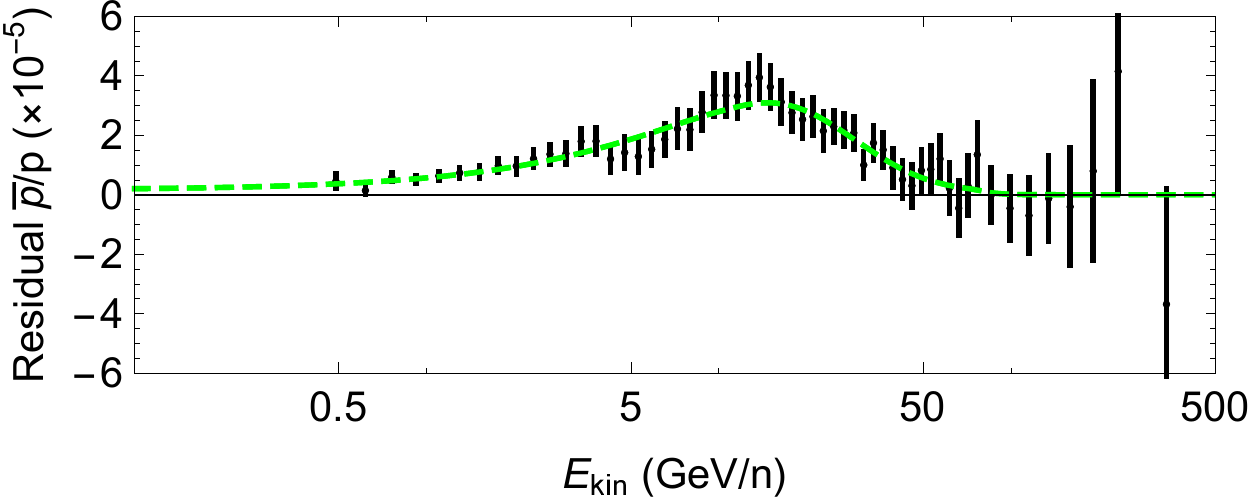}
\includegraphics[width=2.246in,angle=0]{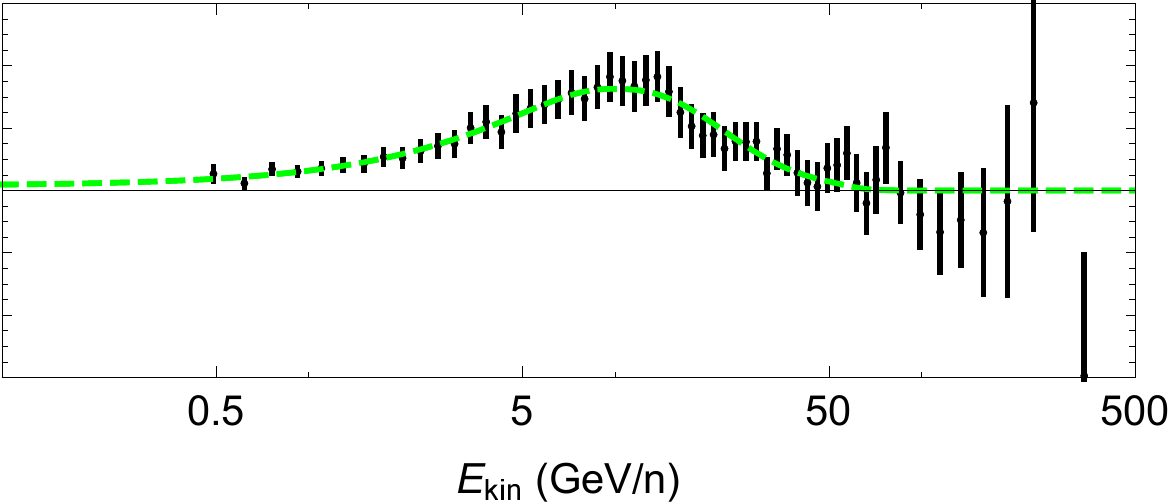}
\includegraphics[width=2.246in,angle=0]{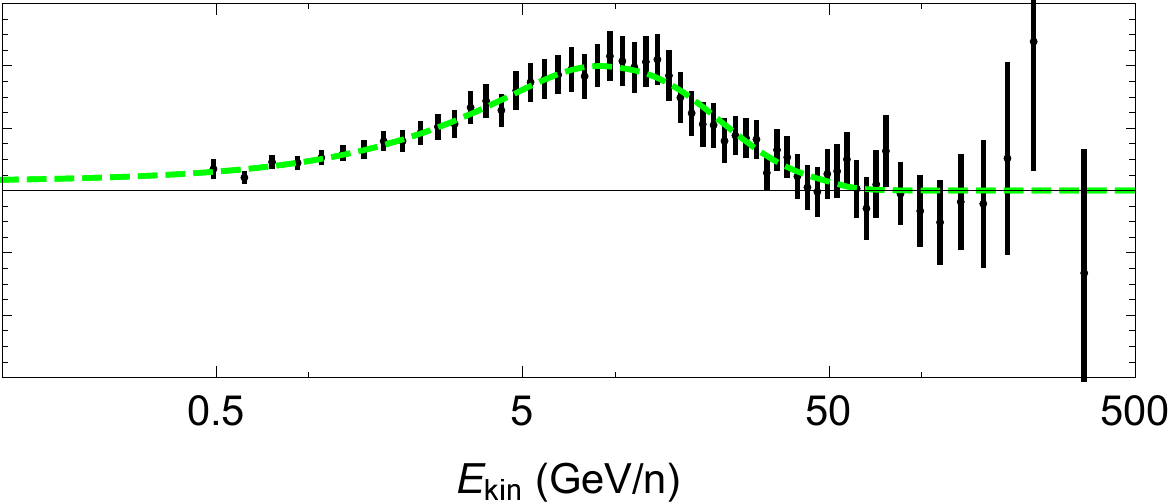}\\
\vskip -0.068in
\caption{As in Figs.~\ref{fig:pbar_DM_NoSAS} and~\ref{fig:pbar_DM_WithSASBC}, but including a contribution from stochastically accelerated secondary antiprotons with a freely floating value of $n_{\rm gas}$}
\label{fig:pbar_DM_WithSAS}
\end{figure*}

\begin{table*}[t]
    \begin{tabular}{cccccccccc}
         \hline
           ISM Model & $K_B$ & $n_{\rm gas}\,({\rm cm}^{-3})$ &  $m_{\chi}\, ({\rm GeV})$ && $\sigma v \, ({\rm cm}^3/{\rm s})$ & Statistical Preference  \\
            \hline \hline
            I & -- & -- &$78.3\pm 4.9 $ && $(1.18 \pm 0.18)\times 10^{-26}$ & 4.7$\sigma$ \\      
            II & -- &-- & $71.0 \pm 3.3$ && $(2.37 \pm 0.28) \times 10^{-26}$ & 7.7$\sigma$  \\
            III & -- & -- & $74.7 \pm 3.8$ && $(4.17 \pm 0.53) \times 10^{-26}$  & 7.4$\sigma$  \\
            \hline
            I & 3.05 & 2.0 (Fixed) & $81.9\pm 6.1$ &&$(1.08\pm 0.19)\times 10^{-26}$ & 3.3$\sigma$ \\      
            II & 5.2  &2.0 (Fixed) & $57.0 \pm 3.1$ && $(1.63 \pm 0.08)\times 10^{-26}$ & 5.1$\sigma$  \\
            III & 3.7 & 2.0 (Fixed) & $51.9 \pm 2.8 $ && $(3.05 \pm 0.14)\times 10^{-26}$  & 5.6$\sigma$  \\     
            \hline
            I & 6.1 & 0.39 (Float) & $78.1\pm5.5$ && $(1.30\pm0.17)\times 10^{-26}$ & 3.4$\sigma$ \\      
            II & 10.4  & 1.28 (Float) & $61.4 \pm 2.5$ && $(1.67 \pm 0.10)\times 10^{-26}$  & 4.6$\sigma$  \\
            III & 7.4 & 1.57 (Float) & $52.3 \pm 3.3$ && $(3.20 \pm 0.30)\times 10^{-26}$  & 5.1$\sigma$  \\                        
            \hline \hline 
        \end{tabular}
\caption{The values of the dark matter mass and annihilation cross section favored by the \textit{AMS-02} antiproton-to-proton ratio, for the case of annihilations to $b\bar{b}$ and for each of the cosmic-ray injection and transport models listed in Table~\ref{tab:ISMBack}. In the top three rows, we have not included any contribution from the acceleration of secondary antiprotons. In the middle three rows, secondary acceleration is included with values of $K_B$ and $n_{\rm gas}$ chosen to reproduce the observed boron-to-carbon ratio. In the bottom three rows, secondary acceleration is included with a freely floating value of $n_{\rm gas}$. The rightmost column indicates the statistical preference for a contribution from annihilating dark matter in each case. We remind the reader that we have marginalized over the parameters associated with the antiproton production cross section and solar modulation at each point in the fit (see Sec.~\ref{sec:method}) and thus conclude that the excess is statistically significant, even in light of these systematic uncertainties.} 
\label{tab:main}
\end{table*}


\section{Stochastic Acceleration of Secondary Cosmic Rays in Supernova Remnants}
\label{sec:SSA}

In the standard picture, cosmic rays are produced when a supernova 
shockfront expands and sweeps through the ISM, trapping particles within its turbulent 
magnetic field structure long enough for them to be accelerated. These particle species 
can also interact with the dense gas on either side of the shockfront. Cosmic rays undergo inelastic scattering and decay at the following rate:
\begin{equation}
\Gamma_{i}(E_{\rm kin}) =  \sigma^{\rm inelastic}_{i} \, \beta \, c \, n_{\rm gas} + \frac{1}{E_{\rm kin}\,\tau^{dec}_{i}},
\label{eq:CRLossRate}
\end{equation}
where $\sigma^{\rm inelastic}_{i}$ and $\tau^{\rm dec}_i$ are the inelastic scattering cross section and lifetime of cosmic ray species, $i$, and $n_{\rm gas}$ is the number density of gas. If the timescale for acceleration is much shorter than that of inelastic scattering or decay, the secondaries will be efficiently accelerated. Following Refs.~\cite{Blasi:2009hv, Mertsch:2009ph, Cholis:2013lwa,Ahlers:2009ae}, 
we assume Bohm diffusion for the cosmic rays near the shockfront:
\begin{eqnarray}
D_{i}^{\pm}(E) &=& \frac{K_{B}\, r_{L}(E)\,c}{3} \\
&=& 3.3 \times 10^{22} \,K_{B} \, 
\left (\frac{1 \mu G}{B} \right )  \left (\frac{E} {1 \rm GeV} \right ) \, \textrm{Z}_{i}^{-1}   {\rm cm}^{2} \,{\rm s}^{-1}, \nonumber
\label{eq:BohmDiff}
\end{eqnarray}
where $r_{L}$ the Larmor radius of the cosmic rays within the magnetic fields and $K_{B} \simeq (B/\delta B )^{2}$ \cite{Blasi:2009hv} quantifies the turbulent nature of the magnetic fields around the shockfront.

The contribution to the cosmic-ray antiproton spectrum from secondary acceleration depends on the value of $K_B$ as well as the density of gas in the scattering region, $n_{\rm gas}$. Both of these parameters have a similar impact on the resulting antiproton spectrum, with larger values leading to a higher antiproton-to-proton ratio at high energies. Increasing $K_B$ or $n_{\rm gas}$ will also increase the boron-to-carbon ratio at high energies, and this information can be used to independently constrain the values of these parameters~\cite{Cholis:2013lwa,Cholis:2017qlb,Tomassetti:2017izg}. 

We begin by adopting values for these parameters that provide a good fit to the observed boron-to-carbon ratio: $n_{\rm gas}=2.0$ cm$^{-3}$ and $K_B=3.05$, 5.2 and 3.7 for ISM models I, II and III, respectively~\cite{Cholis:2013lwa,Tomassetti:2017izg}. The results for these cases are shown in Figs.~\ref{fig:Tim_WithSASBC} and~\ref{fig:pbar_DM_WithSASBC}. The presence of the contribution from accelerated secondaries almost entirely removes the excess at energies above $\sim$100 GeV, erasing the preference for $\sim$1-3 TeV dark matter seen in Fig.~\ref{fig:Tim_NoSAS}~\cite{Cholis:2017qlb}. The evidence for a lighter dark matter particle persists in the presence of accelerated secondaries, however, favoring $m_{\chi}=46-94$ GeV and $\sigma v = (0.7-3.3) \times 10^{-26}$ cm$^3$/s with a statistical significance of 3.3$\sigma$ (see Table~\ref{tab:main}).

By fixing the values of $K_B$ and $n_{\rm gas}$ in our calculations to those which reproduce the observed boron-to-carbon ratio, we are implicitly making the assumption that carbon and protons are accelerated in the same supernova remnants. It is conceivable that carbon nuclei and protons are preferentially accelerated in different subsets of the supernova remnant population, with different average values of $K_B$ and $n_{\rm gas}$. With this possibility in mind, we repeat the above fit, allowing the impact of secondary acceleration to vary. In particular, we set $K_B=6.1$, 10.4 and 7.4 for ISM models I, II and III, respectively, and allow the value of $n_{\rm gas}$ to float freely in the fit (see Table IV of Ref.~\cite{Cholis:2017qlb}). We show the results of this fit in Figs.~\ref{fig:Tim_WithSAS} and~\ref{fig:pbar_DM_WithSAS}. Although the value of $n_{\rm gas}$  takes on different values throughout the mass-cross section plane, we note that our best-fit points correspond to $n_{\rm gas} = 0.39$ cm$^{-3}$, 1.28 cm$^{-3}$ and 1.57 cm$^{-3}$ for ISM models I, II and III, respectively. The fact that these values are similar to those favored to explain the measured boron-to-carbon ratio suggests that carbon nuclei and protons are likely accelerated in the same class of astrophysical sources.

Somewhat surprisingly, the favored parameter space in the right frame of Fig.~\ref{fig:Tim_WithSAS} is entirely surrounded by an excluded region. This stems from the competing demands of the fitting algorithm to match both the (1) high-energy ($\gsim$100 GeV) antiproton excess, which requires large values of $n_{\rm gas}$, and the (2) low-energy ($\sim$10-20 GeV) antiproton excess, which is best fit by dark matter in a scenario with a smaller value of $n_{\rm gas}$. For dark matter masses near our best fit value, moderate values of the annihilation cross section are disfavored, because they force the fit to overproduce the low-energy antiproton excess for the values of $n_{\rm gas}$ that provide the best fit to the high-energy data.

\begin{figure*}[t]
\hspace{-0.0in}
\includegraphics[width=3.45in,angle=0]{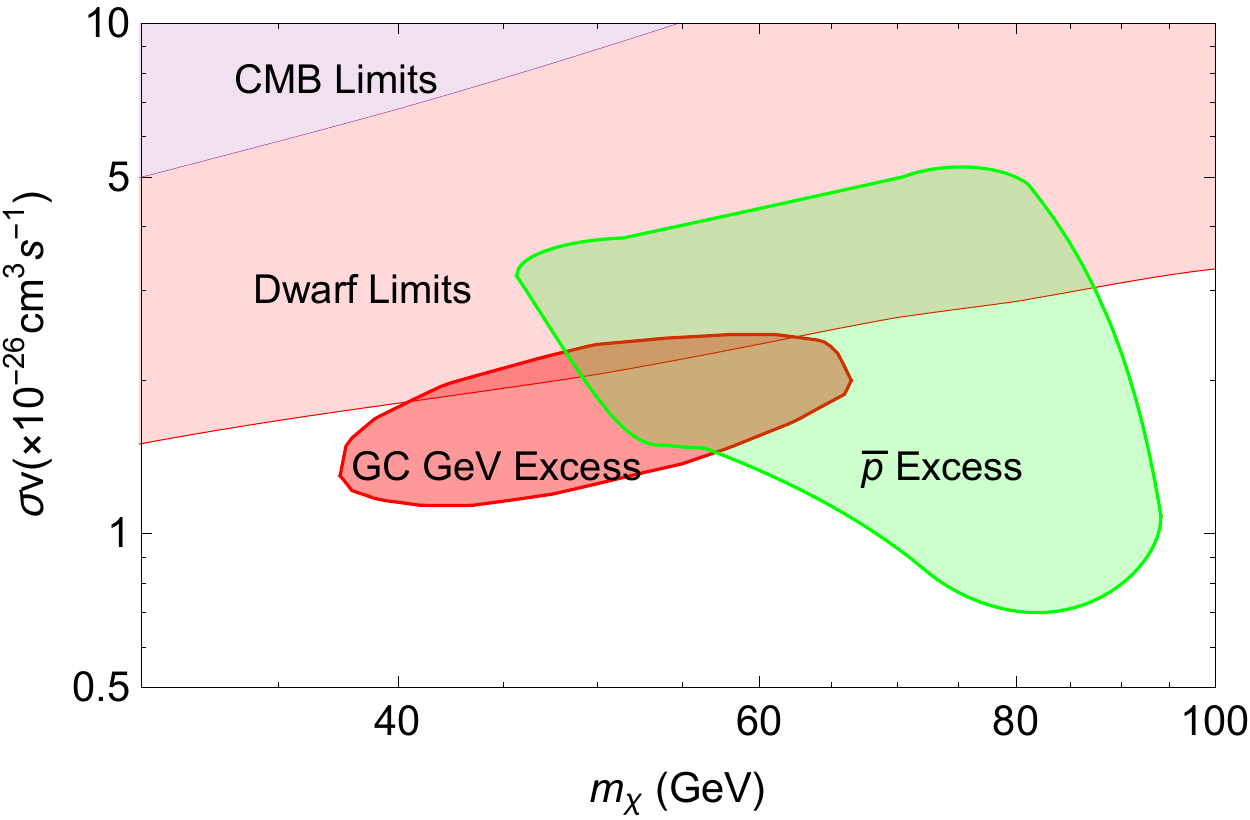}
\includegraphics[width=3.45in,angle=0]{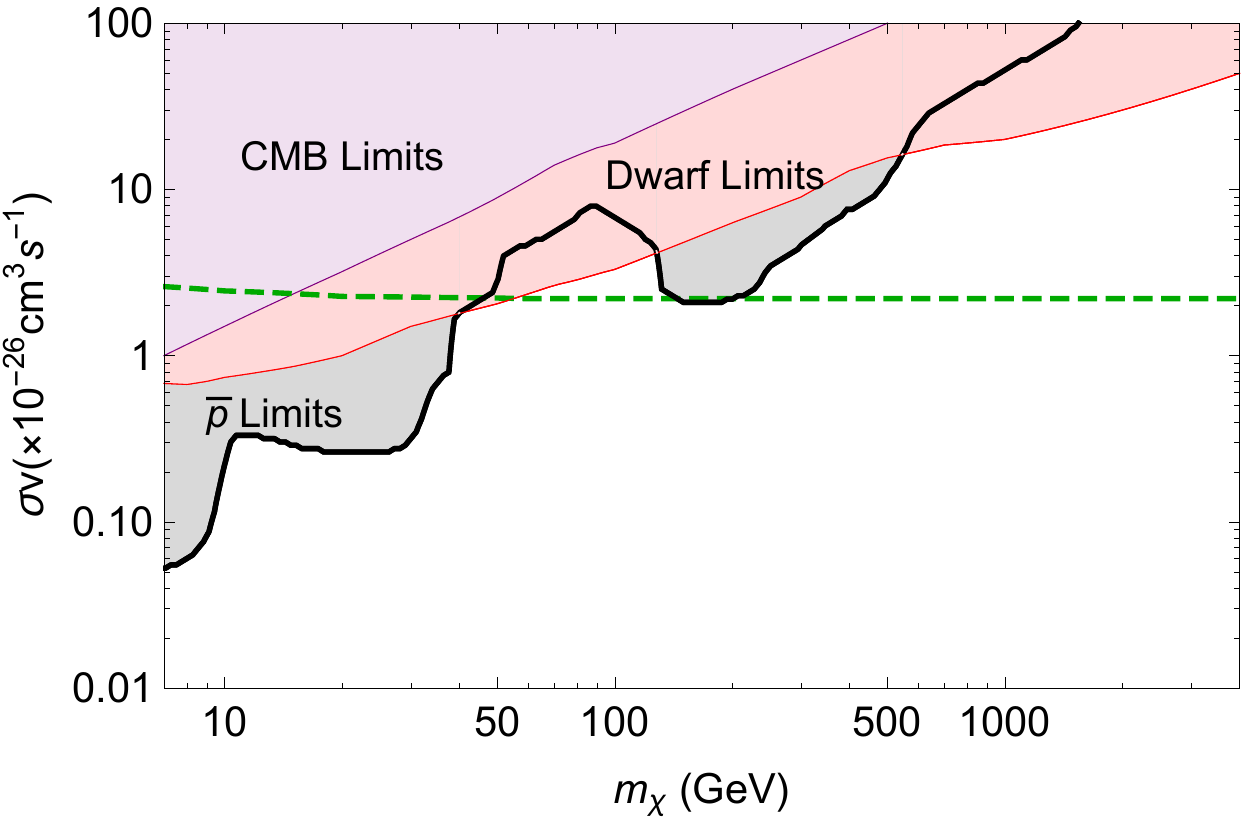}
\vskip -0.1in
\caption{Left frame: The regions of dark matter parameter space favored (within 2$\sigma$) by the \textit{AMS-02} antiproton spectrum (green closed) and the Galactic Center gamma-ray excess (red closed)~\cite{Calore:2014xka}, for the case of annihilations to $b\bar{b}$. Right frame: The upper limit on the dark matter's annihilation section derived from the cosmic-ray antiproton spectrum. Also shown in each frame are the regions excluded by measurements of the cosmic microwave background (purple)~\cite{Aghanim:2018eyx} and by gamma-ray observations of dwarf galaxies (red)~\cite{Fermi-LAT:2016uux}. The dashed green curve denotes the annihilation cross section predicted for dark matter in the form of a simple ($s-$wave) thermal relic.}
\label{fig:DM_limits_andExcess}
\end{figure*}

\section{Discussion and Summary}
\label{sec:conclusions}

In this article, we have studied the cosmic ray antiproton-to-proton ratio measured by {\textit AMS-02}~\cite{Aguilar:2016kjl}, and considered the implications of this measurement for dark matter annihilating in the halo of the Milky Way. Our main results are summarized in Table~\ref{tab:main} and in Fig.~\ref{fig:DM_limits_andExcess}. In each case considered, we find a significant excess of $\sim$10-20 GeV antiprotons, even after marginalizing over a generous range of parameters associated with the effects of solar modulation and the antiproton production cross section. This excess is well fit by annihilating dark matter particles, with a mass and cross section in the range of $m_{\chi}\approx 46-94$ GeV and $\sigma v \approx (0.7-5.2)\times 10^{-26}$ cm$^3/$s, respectively (for the representative case of annihilations to $b\bar{b}$). Other annihilation channels can also provide a good fit, although for slightly different parameter ranges (see Appendix).

Although this result is interesting in its own right, it is particularly intriguing that the range of dark matter models that can accommodate the antiproton excess is very similar to those which could generate the excess of GeV-scale gamma rays observed from the Galactic Center~\cite{Hooper:2010mq,Hooper:2011ti, Abazajian:2012pn, Gordon:2013vta, Daylan:2014rsa,Calore:2014xka,TheFermi-LAT:2015kwa}. In the left frame of Fig.~\ref{fig:DM_limits_andExcess} we compare the regions of dark matter parameter space that are able to account for the gamma-ray excess~\cite{Calore:2014xka} to those favored by the analysis of the antiproton spectrum presented in this study. These two regions overlap, and collectively favor dark matter particles with $m_{\chi} =48-67$ GeV and $\sigma v = (1.4-2.4)\times 10^{-26}$ cm$^3/$s.

Putting the antiproton excess aside for a moment, our analysis also yields stringent constraints on the dark matter annihilation cross section, in many cases competitive with, or more stringent than, other bounds. In the right frame of Fig.~\ref{fig:DM_limits_andExcess}, we show our overall constraint on the dark matter annihilation cross section, which we take to be the weakest of the constraints shown in Figs.~\ref{fig:Tim_NoSAS},~\ref{fig:Tim_WithSASBC} and~\ref{fig:Tim_WithSAS}, evaluated at each value of $m_{\chi}$. Compared to the constraints derived from gamma-ray observations of dwarf spheroidal galaxies~\cite{Fermi-LAT:2016uux}, we find that the limit presented in this study is stronger for dark matter particles with a mass below 40 GeV or between 130 and 540 GeV (for annihilations to $b\bar{b}$).

As this article was being finalized, Ref.~\cite{Cuoco:2019kuu} appeared on the arXiv which addresses many of the same questions discussed here. The authors of Ref.~\cite{Cuoco:2019kuu} reach conclusions that are very similar to our own.

\begin{acknowledgments}

We would like to thank Simeon Bird and Marc Kamionkowski for valuable discussions. 
DH is supported by the US Department of Energy under contract DE-FG02-13ER41958. Fermilab is operated by Fermi Research 
Alliance, LLC, under Contract No. DE- AC02-07CH11359 with the US Department of 
Energy.

\end{acknowledgments}

\bibliography{DM_AMS02_antiprotons}    

\begin{appendix}                  

\section{Results For Other Dark Matter Annihilation Channels}
\begin{figure*}[h]
\hspace{-0.2176in}
\includegraphics[width=2.38in,angle=0]{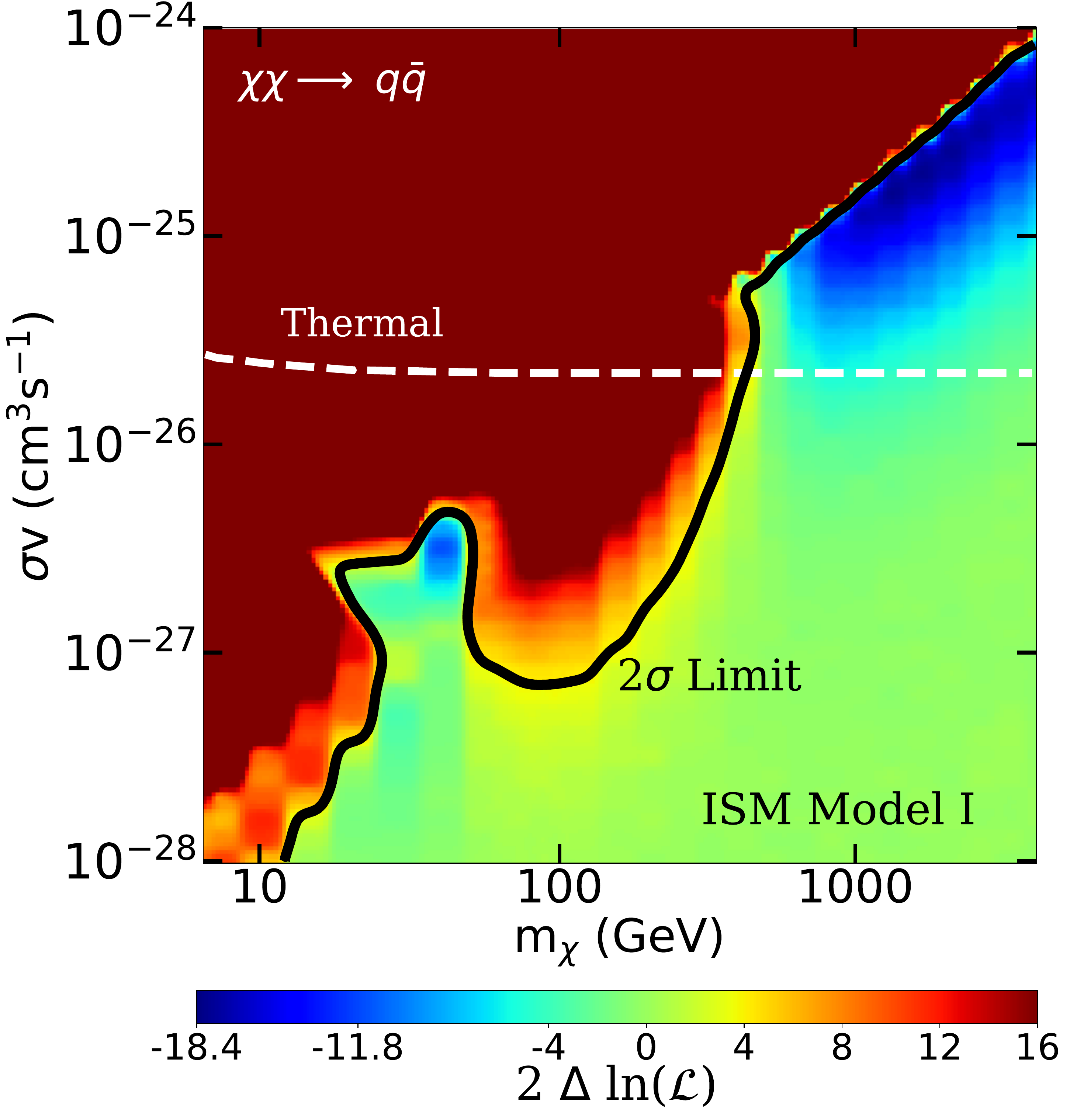} 
\includegraphics[width=2.38in,angle=0]{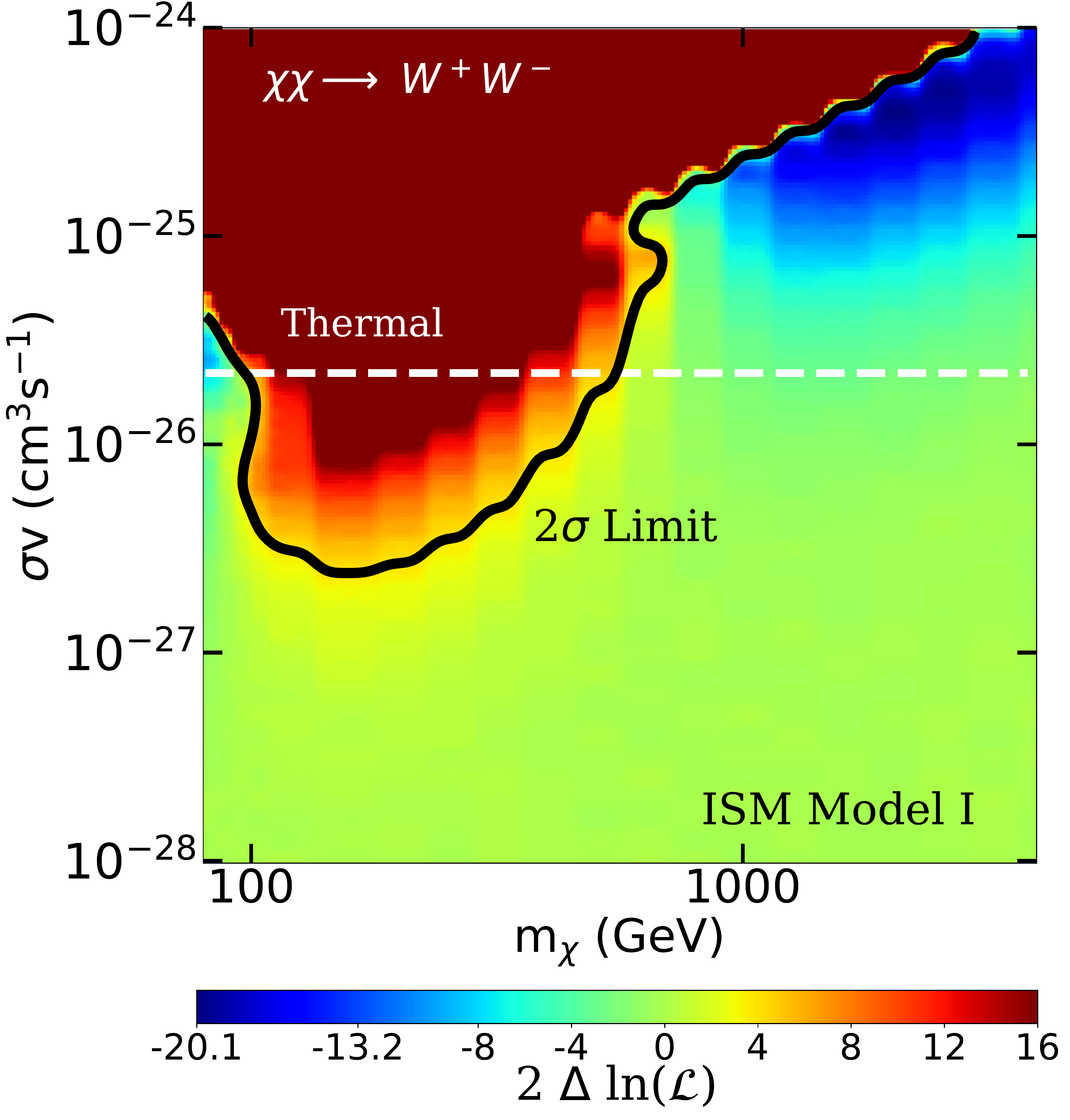}		
\vskip -0.068in
\caption{As in Fig.~\ref{fig:Tim_NoSAS}, but for dark matter that annihilates to light quarks ($u\bar{u}$, $d\bar{d}$) or $W^+ W^-$, and for the case of ISM Model I.}
\label{fig:Tim_channels}
\end{figure*}

\begin{table}[t]
    \begin{tabular}{ccccccc}
         \hline
           Channel &$m_{\chi}$(GeV) & $\sigma v \, ({\rm cm}^3/{\rm s})$ & Statistical Preference  \\
            \hline \hline
            $b\bar{b}$ &  $78.3\pm 4.9 $ & $(1.18 \pm 0.18)\times 10^{-26}$ & 4.7$\sigma$ \\    
            $u\bar{u}$, $d\bar{d}$ &  $38.0 \pm 6.6$ & $(3.12 \pm 0.96)\times 10^{-27}$ & 3.3$\sigma$  \\
            $W^+ W^-$ &  $84.1 \pm 2.2$ & $(3.02 \pm 0.42)\times 10^{-26}$  & 3.6$\sigma$  \\
            \hline \hline 
        \end{tabular}
\caption{As in Table~\ref{tab:main}, but for dark matter that annihilates to $b\bar{b}$, light quarks ($u\bar{u}$, $d\bar{d}$) or $W^+ W^-$, and for the case of ISM Model I.} 
\label{tab:DM_channels}
\end{table}
Throughout this study, we have focused on the representative case of dark matter particles that annihilate to $b\bar{b}$. Clearly this is not the only possibility, and dark matter that annihilates to other final states could also be responsible for the antiproton excess observed by \textit{AMS-02}. In Fig.~\ref{fig:Tim_channels} and Table~\ref{tab:DM_channels} we show our results for dark matter candidates that annihilate to light quarks or to $W^+ W^-$. We also note that hidden sector dark matter candidates could produce a similar spectrum of antiprotons, in particular within the context of Higgs portal models (see Fig.~12 of Ref.~\cite{Escudero:2017yia}). In such a model, the dark matter would annihilate to other hidden sector states, which then decay through mixing with the Standard Model Higgs boson. We note that we utilized~\texttt{PYTHIA}~\cite{Sjostrand:2007gs} for dark matter particles lighter than 86 GeV in the $W^+ W^-$ case, as PPPC~\cite{Cirelli:2010xx} does not generate reliable results in the case of $m_{\chi} \approx m_W$. 

\vspace{-1.0in}
                  
\end{appendix}              

\end{document}